\documentclass{IEEEtran}
\usepackage{amsmath,epsfig,graphicx,subfigure}
\usepackage{booktabs}
\usepackage{array}
\usepackage{multirow}
\usepackage{algorithmicx,algorithm}
\usepackage[noend]{algpseudocode}
\usepackage{caption}
\usepackage{color, soul}
\usepackage[colorlinks,
			linkcolor=black, 
            anchorcolor=black,
            citecolor=black,
            urlcolor=black
            ]{hyperref}

\newcommand{\Region}{\mathcal{L}}
\newcommand{\Videos}{\mathcal{V}}
\newcommand{\Requests}{\mathbf{q}}
\newcommand{\Infra}{\mathbf{a}}

\begin{document}

\title{Understanding Performance of Edge Content Caching for Mobile Video Streaming}

\author{Ge~Ma,
        Zhi~Wang,~\IEEEmembership{Member,~IEEE,}
        Miao~Zhang,
        Jiahui~Ye,\\
        Minghua~Chen,~\IEEEmembership{Senior Member,~IEEE,}
        and~Wenwu~Zhu,~\IEEEmembership{Fellow,~IEEE}
      
\thanks{G. Ma and J. Ye are with Tsinghua-Berkeley Shenzhen Institue, Tsinghua University, Shenzhen 518055, China (e-mail: \href{mailto:mg15@mails.tsinghua.edu.cn}{mg15@mails.tsinghua.edu.cn}; \href{mailto:yejh16@mails.tsinghua.edu.cn}{yejh16@mails.tsinghua.edu.cn}).}

\thanks{Z. Wang and M. Zhang are with the Graduate School at Shenzhen, Tsinghua
University, Shenzhen 518055, China (e-mail: \href{mailto:wangzhi@sz.tsinghua.edu.cn}{wangzhi@sz.tsinghua.edu.cn}; 
\href{mailto:zhangmia15@mails.tsinghua.edu.cn}{zhangmia15@mails.tsinghua.edu.cn}).}

\thanks{M. Chen is  with the Department of Information Engineering, The Chinese
University of Hong Kong, Hong Kong (e-mail: 
\href{mailto:minghua@ie.cuhk.edu.hk}{minghua@ie.cuhk.edu.hk}).}

\thanks{W. Zhu is with the Department of Computer Science and Technology, Tsinghua-Berkeley Shenzhen Institue, Tsinghua National Laboratory  for Information Science and Technology, 
Beijing Key Laboratory of Networked Multimedia, Tsinghua University, Beijing 100084, China (e-mail: \href{mailto:wwzhu@tsinghua.edu.cn}{wwzhu@tsinghua.edu.cn}).}
}

\maketitle

\begin{abstract} 	

Today's Internet has witnessed an increase in the popularity of mobile video streaming, which is expected to exceed $3/4$ of the global mobile data traffic by 2019. To satisfy the considerable amount of mobile video requests, video service providers have been pushing their content delivery infrastructure to edge networks--from regional CDN servers to peer CDN servers (e.g., smartrouters in users' homes)--to cache content and serve users with storage and network resources nearby. Among the edge network content caching paradigms, Wi-Fi access point caching and cellular base station caching have become two mainstream solutions. Thus, understanding the effectiveness and performance of these solutions for large-scale mobile video delivery is important. 
However, the characteristics and request patterns of mobile video streaming are unclear in practical wireless network.
In this paper, we use real-world datasets containing $50$ million trace items of nearly $2$ million users viewing more than $0.3$ million unique videos using mobile devices in a metropolis in China over $2$ weeks, not only to understand the request patterns and user behaviors in mobile video streaming, but also to evaluate the effectiveness of Wi-Fi and cellular-based edge content caching solutions. 
To understand performance of edge content caching for mobile video streaming, we first present \emph{temporal} and \emph{spatial} video request patterns, and we analyze their impacts on caching performance using frequency-domain and entropy analysis approaches. We then study the behaviors of mobile video users, including their mobility and geographical migration behaviors, which determine the request patterns. Using trace-driven experiments, we compare strategies for edge content caching including LRU and LFU, in terms of supporting mobile video requests. We reveal that content, location and mobility factors all affect edge content caching performance.
Moreover, we design an efficient caching strategy based on the measurement insights and experimentally evaluate its performance. The results show that our design significantly improves the cache hit rate by up to $30\%$ compared with LRU/LFU.

\end{abstract}


\section{Introduction} \label{sec:intro}

Global mobile video traffic reached $3.7$ EB/month at the end of 2015, and it is predicted that over $3/4$ of the global mobile data traffic will be video traffic by 2019 \cite{cisco}. This trend is further accelerated by the rapid growth of online/mobile social media and mobile networks: video clips are increasingly being generated by users and instantly shared with their friends. In contrast to conventional live and on-demand video streaming that are consumed using TVs and PCs, mobile video streaming is generally watched by users on mobile devices with wireless connections, i.e., 3G/4G cellular or Wi-Fi. User behaviors and wireless network quality in mobile video streaming \cite{hu2014quality,liang2015wireless} can be quite different from those in conventional video streaming \cite{adhikari2012unreeling,mukerjee2015enabling}, thus requiring improvements in the delivery of mobile video streaming.

To meet the sky-rocketing increase in bandwidth requirements resulting from data-intensive video streaming and to reduce the monetary cost for renting expensive resources in conventional content delivery networks (CDNs), video service providers are pushing their content delivery infrastructure closer to users to utilize network and storage resources in households for content delivery \cite{baochun-tomccap-streaming2013}, including caching content over femtocells \cite{golrezaei2012femtocaching} and replicating video content via Wi-Fi smartrouters in households. Youku, one of the largest online video providers in China, has deployed over $300$K smartrouters in its users’ homes in less than one year, expecting to transform a large fraction of its users ($250$M) into such content delivery peer nodes \cite{Ming-nossdav16}. To serve users with good quality of experience using the new edge network solutions, it is important to answer the following questions: (1) What are the video request patterns in mobile video streaming, how do users behave in today's mobile video systems, and what is the implication of their behaviors on edge network video content delivery (Sec.~\ref{sec:mobile-video} and Sec.~\ref{sec:userpatterns})? (2) How is the quality of user experience in the mobile video sessions (Sec.~\ref{caching_evaluation})? (3) Can today's mobile network infrastructure appropriately satisfy the mobile video streaming demand (Sec.~\ref{infrastructure})? (4) What strategies can be applied to best support mobile video content delivery (Sec.~\ref{sec:application})?

Several measurement studies have been conducted to address the above questions. However, such measurement studies are challenging because many different factors are involved, including user behaviors (i.e., mobility pattern and video preference), video content characteristics, and mobile network characteristics. Previous studies generally focus on a single aspect, e.g., studying the popularity of mobile video content \cite{brodersen2012youtube,xu2015forecasting}, user mobility behaviors \cite{das2014contextual}, or network strategies to support mobile video streaming, e.g., content replication \cite{gitzenis2013asymptotic}. The limitation of the previous studies is that they have not considered the joint impact of user behaviors, content characteristics and wireless network deployment, on edge network content delivery.

In this paper, we propose to address the above questions from the perspectives of both the mobile video service and wireless network providers. From the perspective of mobile video service, we study how users view mobile videos, including their mobility patterns in video sessions and the content selection in different locations, and we build a mobile video consumption model. From the perspective of wireless network provider, we present how the mobile video requests can be served by both the Wi-Fi and cellular infrastructures that are commonly used by today's users, and we provide insights on how to improve the QoS of wireless networks according to their video request patterns.

Our contributions are summarized as follows:

First, we use large-scale datasets to study user behaviors in a real-world mobile video streaming system, covering $50$ million sessions of nearly $2$ million users viewing more than $0.3$ million videos using mobile devices in $2$ weeks. Using frequency-domain and entropy analyses \cite{brodersen2012youtube,wang2015understanding}, we show that mobile video requests exhibit unique spatial and temporal patterns that can significantly affect the performance of content caching strategies in edge networks. (1) We observe a skewed geographic request distribution in the mobile video system, and the number of requests is highly affected by the regular mobility patterns of users. For example, the number of requests in train stations is much larger than that in residential areas. (2) We observe that videos with lower popularity have more uniform distribution of requesting locations, while videos with higher popularity have more skewed distribution. Surprisingly, the increase of multi-location users (who request videos in different locations, i.e., mobile users) in a location does not increase the requested number of unique videos, which is different from conventional single-location users (who request videos only in the same locations, i.e., home users). (3) In the frequency-domain analysis, we observe that the number of requests in locations with different functionalities over time has $3$ major periods, e.g., residential areas have an obvious period of $8$ hours, which can be used to predict the future traffic in content delivery.

Second, we further investigate how user behaviors determine the above request patterns. We reveal that both mobility and geographic migration behaviors of users can significantly affect mobile video requests. In particular, we show that the mobility behaviors of users are heterogeneous, e.g., a number of multi-location users request videos intensively and request them in different locations, whereas there is a large fraction of users who only request a small number of videos in the same location. For the geographic migration behaviors, we observe that (1) users have regular commute behaviors, involving $2$--$3$ regularly visited locations where they tend to request mobile videos, and (2) it is common for users to move between the same type of locations (e.g., residential) and issue video requests. \emph{These observations suggest that joint caching strategies over multiple locations can improve the caching performance}. 

Third, we compare the effectiveness of Wi-Fi and cellular-based edge network caching solutions, and we discuss the potential improvement on mobile video streaming to today's wireless networks. Based on our edge network traces covering $1,055,881$ Wi-Fi APs and $69,210$ cellular base stations, we investigate conventional caching strategies, including least recently used (LRU) and least frequently used (LFU) for edge network mobile video delivery. We first show that most of today's Wi-Fi and cellular deployments are close enough to the mobile requests of users in different locations; however, although Wi-Fi and cellular have different deployment strategies, they cannot well serve different categories of mobile video users. Second, we show that a number of factors including user mobility, content popularity, cache capacity, and caching strategies affect the caching performance for both Wi-Fi and cellular caching for mobile video delivery. For example, unpopular videos attract users mostly from few locations where users have particular interests in the content, and caching strategies have various influences on different categories of users.

Finally, motivated by the measurement insights, we design a geo-collaborative caching strategy for mobile video delivery, which jointly considers mobile video request patterns, user behaviors and the deployment of wireless networks. Based on real-world trace-driven experiments, we show that our design achieves a $20\%$ (resp. $30\%$) cache hit rate improvement and a $20\%$ (resp. $30\%$) service rate improvement compared with conventional LRU (resp. LFU) caching strategies.

The remainder of this paper is organized as follows. We discuss the related works in Sec. \ref{sec:relatedwork}. We present the datasets used in our measurement studies in Sec.~\ref{sec:measure}. We study the temporal and spatial request patterns and the content characteristics in a mobile video system in Sec.~\ref{sec:mobile-video}. 
We measure user behaviors in mobile video streaming sessions and how they affect the quality of mobile streaming in Sec.~\ref{sec:userpatterns}. We compare the effectiveness of Wi-Fi and cellular-based edge content delivery solutions and discuss the potential improvement to today's wireless networks to improve mobile video streaming in Sec.~\ref{sec:edgenet}. We present the details of our caching strategy and evaluate its performance in Sec.~\ref{sec:application}. Finally, we conclude the paper in Sec. \ref{sec:conclusion}.

\section{Related Work} \label{sec:relatedwork}

There are four main research areas related to our work: video measurement, user mobility behaviors, edge video delivery and edge network caching strategies. 

\subsection{Video Measurement}
There are several prior studies that focus on the properties of videos and how to model and predict the popularity of such videos. One of the works investigates the relationship between the popularity and location of online videos \cite{brodersen2012youtube,li2014geographic}. This work finds that videos exhibit a geographical distribution of interest, with users arising from a confined and single area rather than from a global area, and it provides new insights on how the geographic reach of a video changes as its popularity peak and then fades away. 
The prediction of video popularity has also been studied based on historical information given by early popularity measures \cite{xu2015forecasting,pinto2013using}. Two novel models are proposed, which are able to better distinguish between videos with different popularities, by assigning different weights to samples with different popularities and exploring the similarity between the video and known samples within the monitoring period. Our study on mobile video differs from these works since our analysis focuses not only on time period (hour level), but also on entropy analysis. In addition, the geographic locations that we measured are more specific, allowing us to obtain a comprehensive relationship between temporal and spatial video request patterns.

\subsection{User Mobility Behaviors}
There are also several prior studies that focus on characterizing mobile video traffic. Li et al.~\cite{Li2012Watching} focus on analyzing the main discrepancies when users access video-on-demand systems using either Wi-Fi or 3G connections. They study the factors that affect mobile users' interests and video popularity. Li et al.~\cite{li2014geographic} characterize the geographical patterns on a large-scale, commercial, mobile video-on-demand system and analyze the temporal evolution trends of the geographical popularity, which reveal distinct behaviors of popular and unpopular videos. However, they only use coarse-grained (in province) location information, which differs from our study in which the latitudes and longitudes of users are analyzed to obtain useful insights about the relationship between user mobility and video request patterns. Recently, Wang et al.~\cite{wang2015understanding} model the mobile traffic patterns of large-scale cellular base stations deployed in a city. Their work contributes some valuable information for Internet service providers, mobile users, and government management of mobile network resources.

With the development of new location-sensing technologies, the information about the locations of users has become available. Toole et al.~\cite{toole2012inferring} use the dynamic data generated by mobile phones to measure spatiotemporal changes in population, and identify the relationship between land use and dynamic population. Considering that sharing precise location information may cause leaks of privacy information, Das et al.~\cite{das2014contextual} study the contextual locations of users by passively monitoring the mobile network traffic of many location-based services, which only rely on contextual location. In contrast to these works, our study focuses on providing an understanding of user mobility and geographical migrations when using mobile video services. A QoE modeling framework with user, system and context components is created for a mobile video environment, taking mobile user, mobile device, mobile network and mobile video service into consideration \cite{song2012understanding}. Thus, users requesting mobile videos may benefit from the model, and video providers could also develop effective strategies to improve the user experience. Furthermore, the viewing conditions of mobile video can be described in terms of three main factors: display size and viewing distance, surrounding luminance, and body movements of the viewer \cite{xue2012study}. It incorporates all three of these important factors into optimizing video coding and delivery for mobile devices. Some studies show that users' cooperation can effectively reduce the servers' burden, such as delay and bandwidth, confirmed to be an attractive solution to limit the costs incurred by content providers \cite{ciullo2015much,ciullo2013asymptotic}.

\subsection{Edge Video Delivery}
The substantial demand for bandwidth from data-intensive applications has challenged the traditional content delivery paradigms: the content delivery network (CDN), including its variations ISP-operated CDN \cite{cho2011can}, content-provider-operated CDN \cite{wang2015cpcdn}, and peer-to-peer CDN \cite{zhang2015unreeling,zhao2013peer}. Because mobile video content has occupied most of the mobile network traffic, caching videos in the network edge (i.e., femtocells or Wi-Fi APs) has become a common solution. Building caches at the network edge is an appealing solution since the cost of network equipment, such as cellular base stations, substantially exceeds the cost of installing a cache \cite{roberts2013exploring}. Furthermore, if videos can be fetched from a local cache rather than CDN servers, the large delays can be significantly reduced \cite{ahlehagh2014video}. Golrezaei et al.~\cite{golrezaei2012femtocaching} envision femtocell-like base stations called helpers, with weak backhaul links but large storage capacity, which can assist in the macro base station by handling requests for popular files that have been cached. Based on a real measurement study of mobile video viewing logs from a leading Internet video provider for $14$ days, Lin et al.~\cite{Lin2013Mobile} study the potential of peer-assisted video delivery in Wi-Fi mobile networks aiming to reduce server load. Moreover, Zhou et al.~\cite{zhou2015video} study how video popularity changes over time and varies among different categories, and they apply the results to design video caching strategies in CDN servers. 

\subsection{Edge Network Caching Strategies}

The impact of content popularity dynamics on cache performance can be captured by an analytical model under the assumption that requests at different caches are independent \cite{garetto2015efficient}. Based on this assumption, a threshold-based caching scheme is proposed for wireless access networks, which replicates content that is requested more times than the given threshold \cite{leconte2016placing}. To investigate collaborative caching, coded caching strategies for heterogeneous wireless networks have been proposed to balance the cost among base station transmission, access point storage and user connection latency \cite{hachem2015content,maddah2014fundamental}. Distributed caching architectures have also been proposed to replicate content close to users to reduce the average video delivery delay \cite{poularakis2016caching}.

To the best of our knowledge, we are the first to jointly measure both the mobile video request patterns, user mobility behaviors, and the deployment of wireless networks to investigate the performance of wireless network caching and to design an efficient caching strategy for mobile video delivery.

\section{Datasets on Mobile Video Streaming and Edge Networks} \label{sec:measure}

In this section, we present how we collect the datasets used in our study.

\subsection{Mobile Video Behavior Dataset}

The mobile video behavior dataset is collected by a video provider company in China. How users view videos in the mobile video streaming app has been recorded.
The dataset spans $2$ weeks and covers $2$ million users watching $0.3$ million unique videos in Beijing. In each trace item, the following information is recorded: (1) The device identifier, which is unique for different devices and can be used to track users; (2) The timestamp when the user starts to watch the video; (3) The location where the user watches the video: the video player reports the location either collected from the device's built-in GPS function or inferred from the network parameters (e.g., cellular base station); (4) The title of the video, as shown in Table \ref{tab:video}.

\begin{table}[!t]
\renewcommand\arraystretch{1.5}
\small
\begin{center}
\caption{\label{tab:video}Mobile Video Behavior Dataset.}
\begin{tabular}{p{3cm}p{4.5cm}}
\toprule
\multicolumn{1}{c}{Field} & \multicolumn{1}{c}{Description} \\
\midrule
{User ID (anonymized)} & The unique identifier of each user\\
\multirow{2}{*}{Request time} & The specific time that the user requests a video\\
\multirow{2}{*}{Latitude and longitude} & The position of current request is issued\\
\multirow{2}{*}{Video content} & The name and some basic information of the video \\
\bottomrule
\end{tabular}
\end{center}
\end{table}

\subsection{Wi-Fi and Cellular Network Dataset}

We also study how today's edge network content delivery paradigms can be supported by both Wi-Fi and cellular solutions \cite{heoptimizing,xu2011cellular}. 

\textbf{Wi-Fi AP Information.} The Wi-Fi and cellular network dataset is provided by Tencent Wi-Fi \cite{tencent}, a mobile app that asks users to respond to questions on how they use Wi-Fi/cellular networks. In particular, we collected over $1$ million Wi-Fi APs in Beijing city, including the basic service set identifier (BSSID) of Wi-Fi APs and the location of the Wi-Fi hotspots. This valuable dataset samples a large fraction of Wi-Fi APs that are actually deployed in Beijing, allowing us to determine whether these APs can provide content delivery functionality for mobile video streaming. Table \ref{tab:network} shows the details of the dataset: each trace item contains the latitude and longitude of the AP and the point of interest (PoI) information of the AP (e.g., hotel).

\textbf{Cellular Base Station Information.} Our dataset also contains cellular network information, including locations, IDs, and location area code (LAC) of over $70$ thousand cellular base stations.

\begin{table}[!t]
\small
\renewcommand\arraystretch{1.5}
\begin{center}
	\caption{Wi-Fi and Cellular Network Dataset.}\label{tab:network}
	\begin{tabular}{p{3cm}p{4.5cm}}
	\toprule
		\multicolumn{1}{c}{Field} & \multicolumn{1}{c}{Description} \\
	\midrule
		{MAC} & The MAC address of the device\\
		{Latitude and longitude} & The specific position of the device\\
		\multirow{2}{*}{LAC and Cell IDs} & The location code and cell ID of the base station\\
		{MNC ID} & The mobile network code\\
		{Address of AP} & The detailed address of the device \\
		\multirow{2}{*}{PoI} & A functionality description of the location, e.g., \emph{university}\\
	\bottomrule
	\end{tabular}
\end{center}
\end{table}

\section{Request Patterns in Mobile Video Streaming} \label{sec:mobile-video}

In this section, we first investigate the popularity distribution of mobile videos;
we then study the spatial and temporal patterns of users' video requests in mobile video streaming and present how content affects mobile video requests; finally, we present the implications of such request patterns.

\subsection{Popularity Distribution}
\label{sec:pop}

\begin{figure*}[!t]
     \centering
         \subfigure[Popularity distribution]{
             \label{fig:popularity}
             \includegraphics[width=0.32\linewidth]{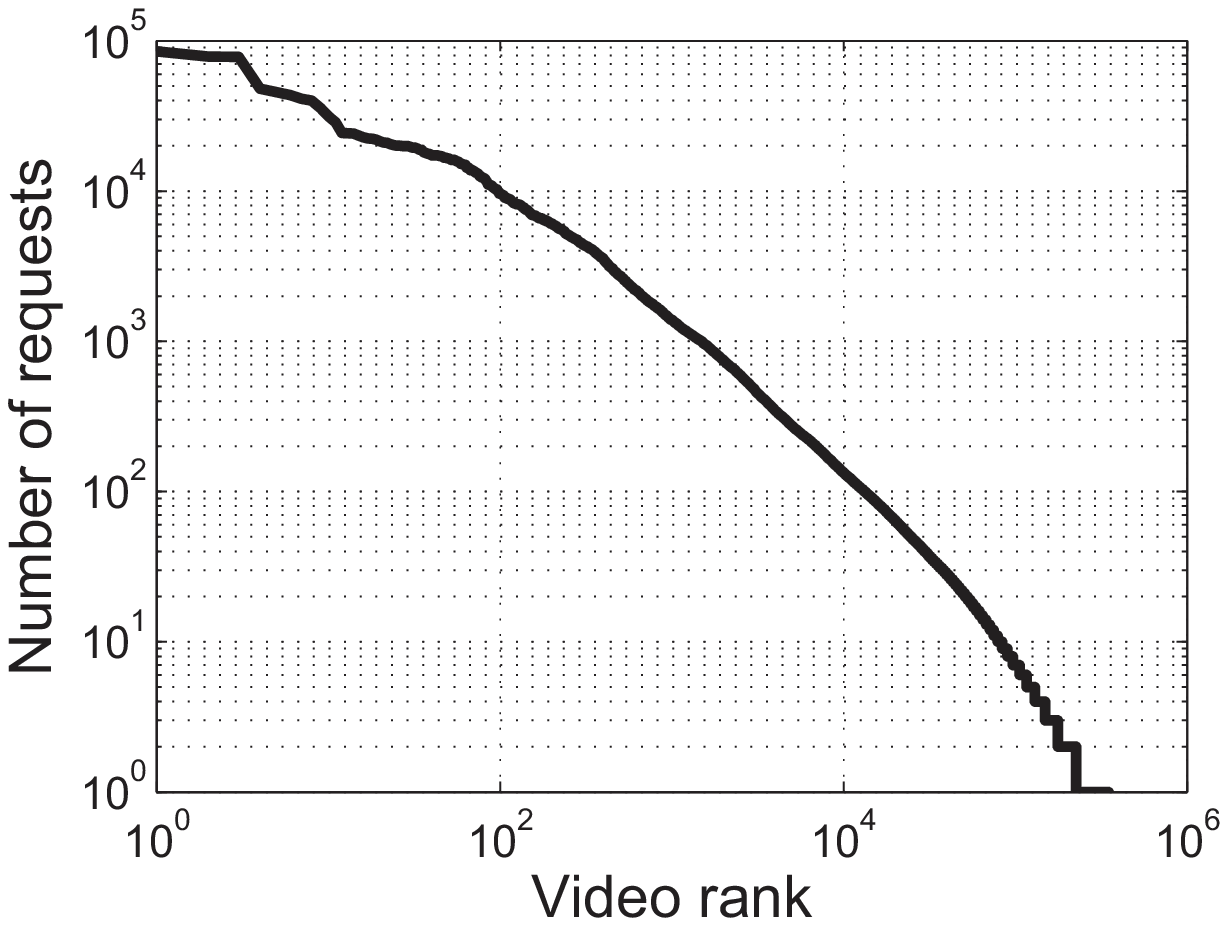}}
         \hfill
        \subfigure[Popularity change over time]{
             \label{fig:popularity_time}
             \includegraphics[width=0.32\linewidth]{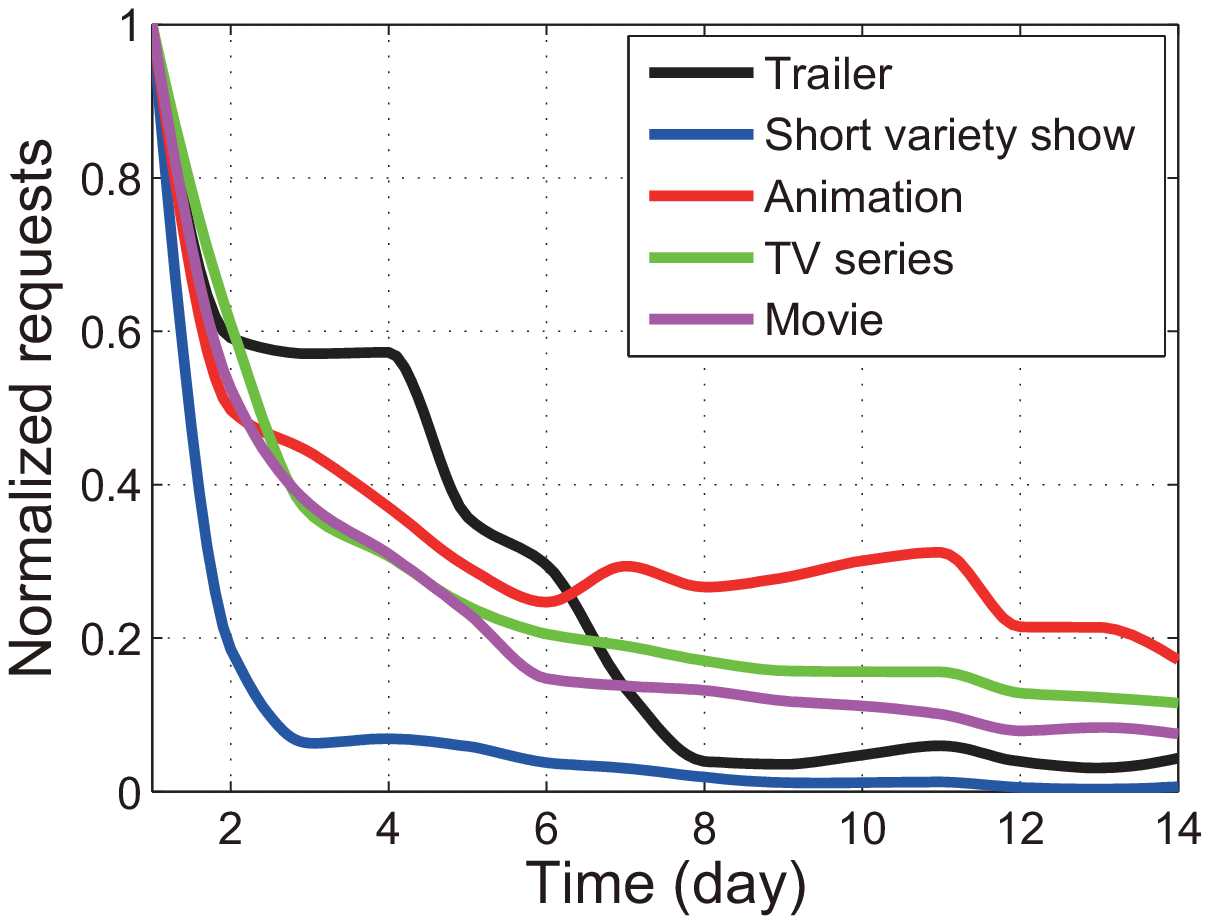}}
         \hfill
         \subfigure[Local popularity of the most popular videos in the whole system]{
             \label{fig:popularity_location}
             \includegraphics[width=0.32\linewidth]{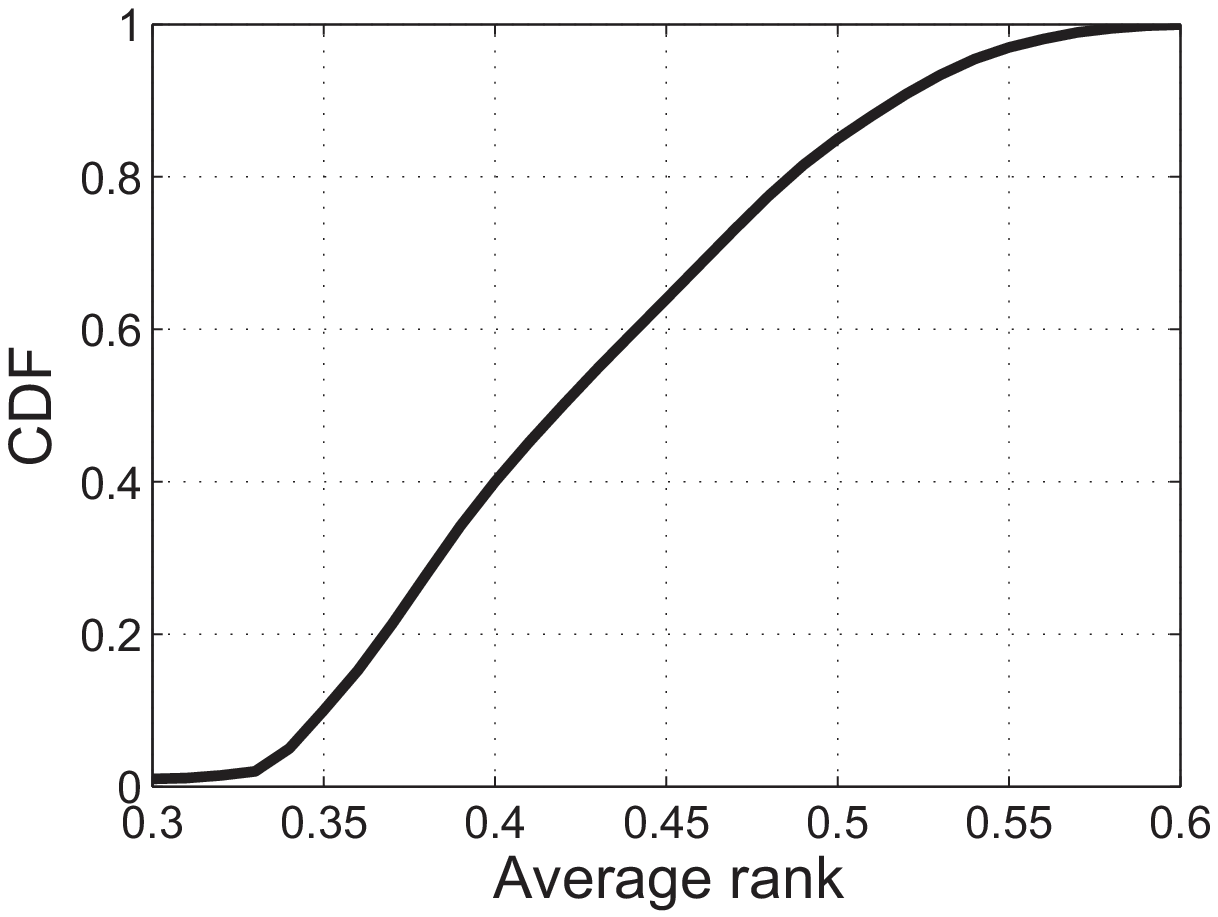}}                 
     \caption{Characteristics of mobile video popularity.}
     \label{fig:popularity1}
\end{figure*}

We first describe the popularity distribution of mobile videos. As illustrated in Fig.~\ref{fig:popularity}, we observe that the popularity of mobile video content also follows a power-law distribution.

Fig.~\ref{fig:popularity_time} shows the average normalized number of daily requests for different video categories over time, for the $1000$ most popular videos. We observe that trailer has the smallest decreasing rate, short variety show has the largest decreasing rate, and the animation category has the longest lifetime.

Furthermore, we investigate the popularity of videos in different locations by studying the popularity rank of the $1,000$ most popular videos (top $0.3\%$ in the entire system). In Fig.~\ref{fig:popularity_location}, we plot the CDF of the average popularity rank of these videos in $1,000$ locations where they are requested. We observe that the top $0.3\%$ videos have quite different popularity ranks in different locations: the average popularity rank for these videos is below the top $40\%$ in as many as $60\%$ of the locations. \emph{This observation indicates that global popularity cannot be directly used to infer the local popularity of mobile video content.} Thus a local caching strategy is more suitable than a global strategy in current mobile video systems.

\subsection{Spatial and Temporal Patterns of Mobile Video Requests}
\label{sec:request_patterns}

To study the mobility patterns of viewers, we assume that the users' requests can be served by the nearest Wi-Fi APs or cellular BSes. Thus, we first classify all the users in the mobile video streaming system into two categories: \emph{multi-location users}, who request videos in different locations (APs/BSes) within \emph{one day} in the traces, and \emph{single-location users}, whose requests are all issued from the same location (APs/BSes) within \emph{one day}. Note that a user may be a multi-location user or a single-location user on different days.

\subsubsection{Skewed Geographical Request Distribution}

\begin{figure*}[t]
     \begin{minipage}[t]{.31\linewidth}
          \centering
                \includegraphics[width=\linewidth]{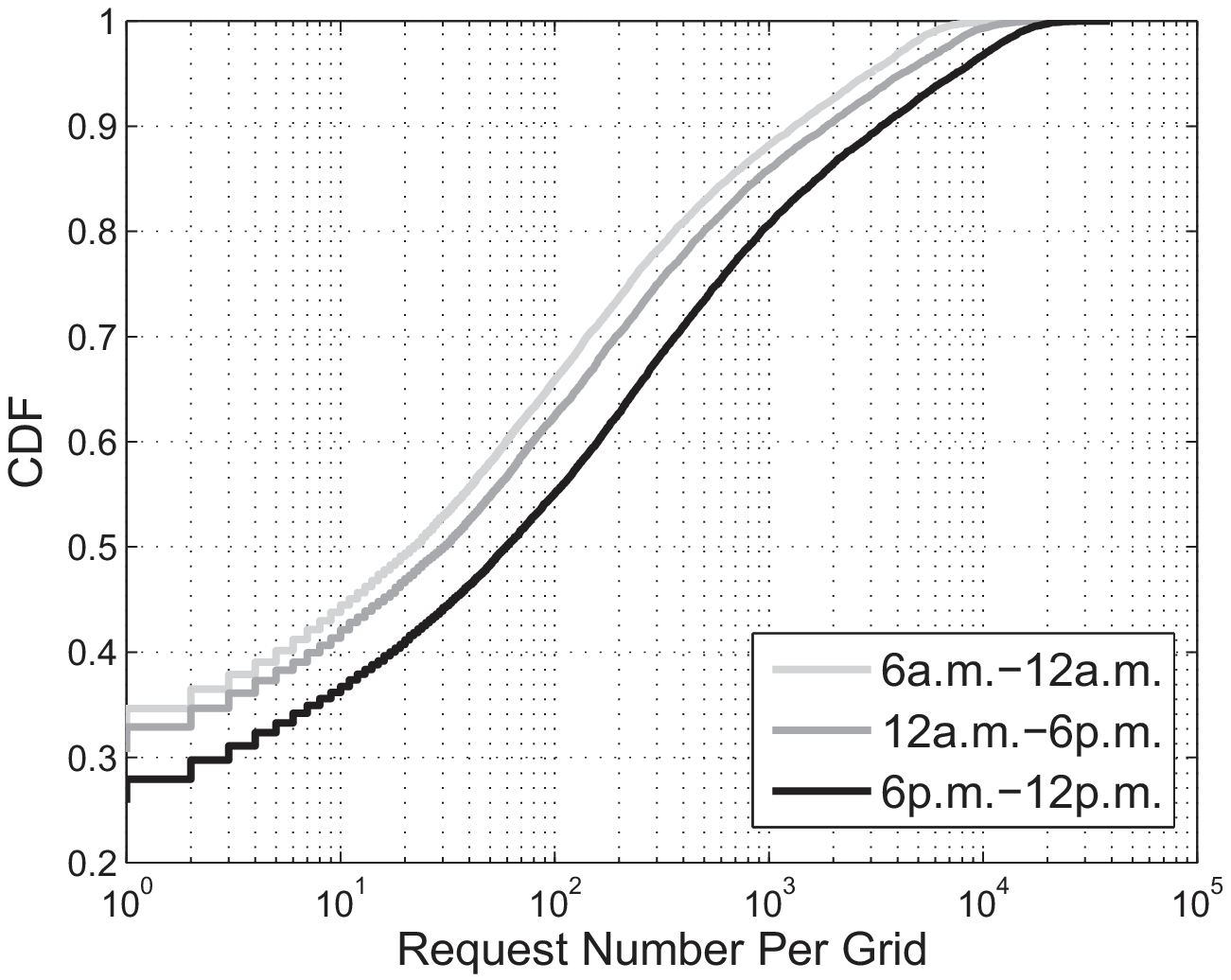}
          \caption{CDF of the number of requests in the locations partitioned geographically.}
          \label{fig:cdf-location-request}
     \end{minipage}
     \hfill
     \begin{minipage}[t]{.31\linewidth}
          \centering
               \includegraphics[width=\linewidth]{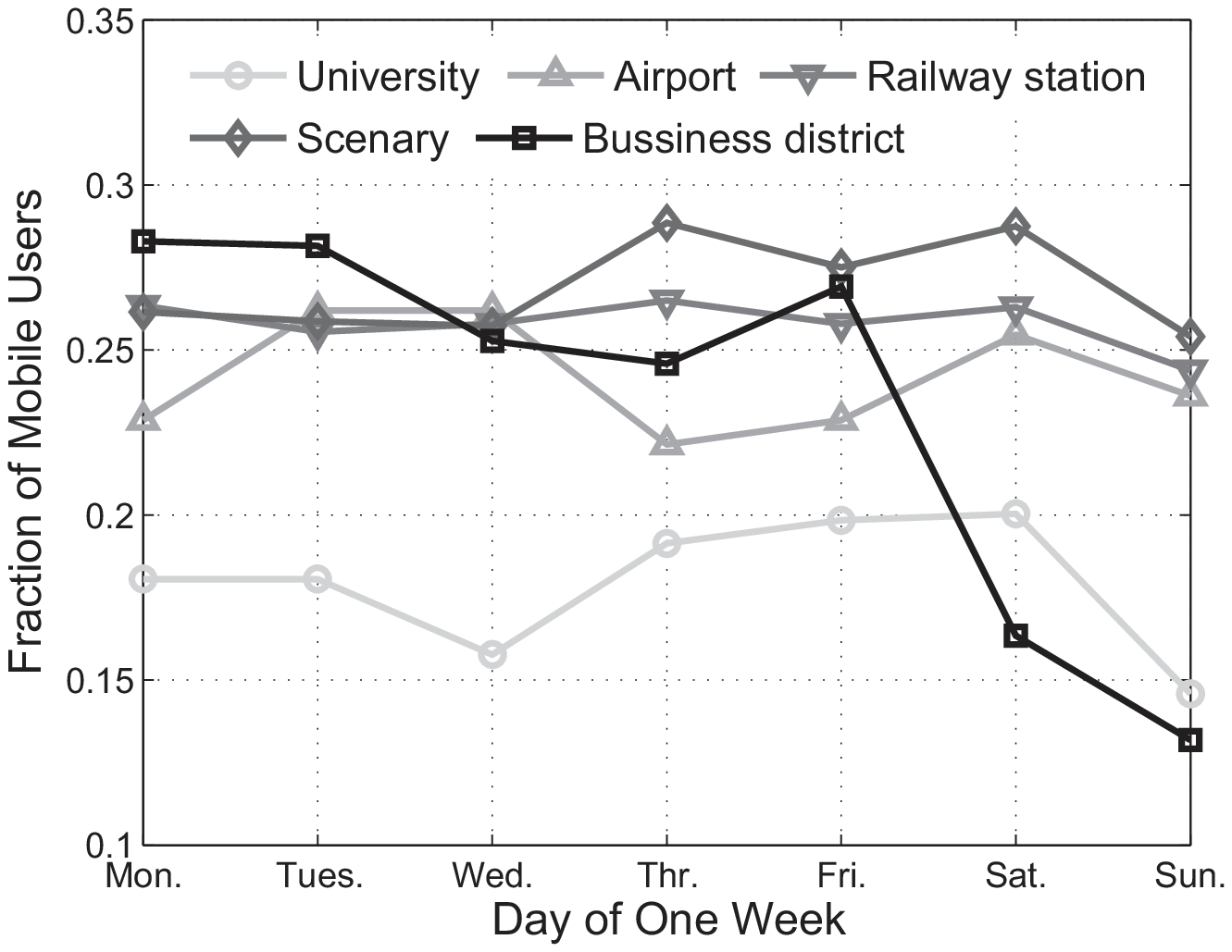}
          \caption{Evolution of multi-location users in different locations.}
          \label{fig:jhmobiuser}
     \end{minipage}
     \hfill
     \begin{minipage}[t]{.31\linewidth}
          \centering
               \includegraphics[width=\linewidth]{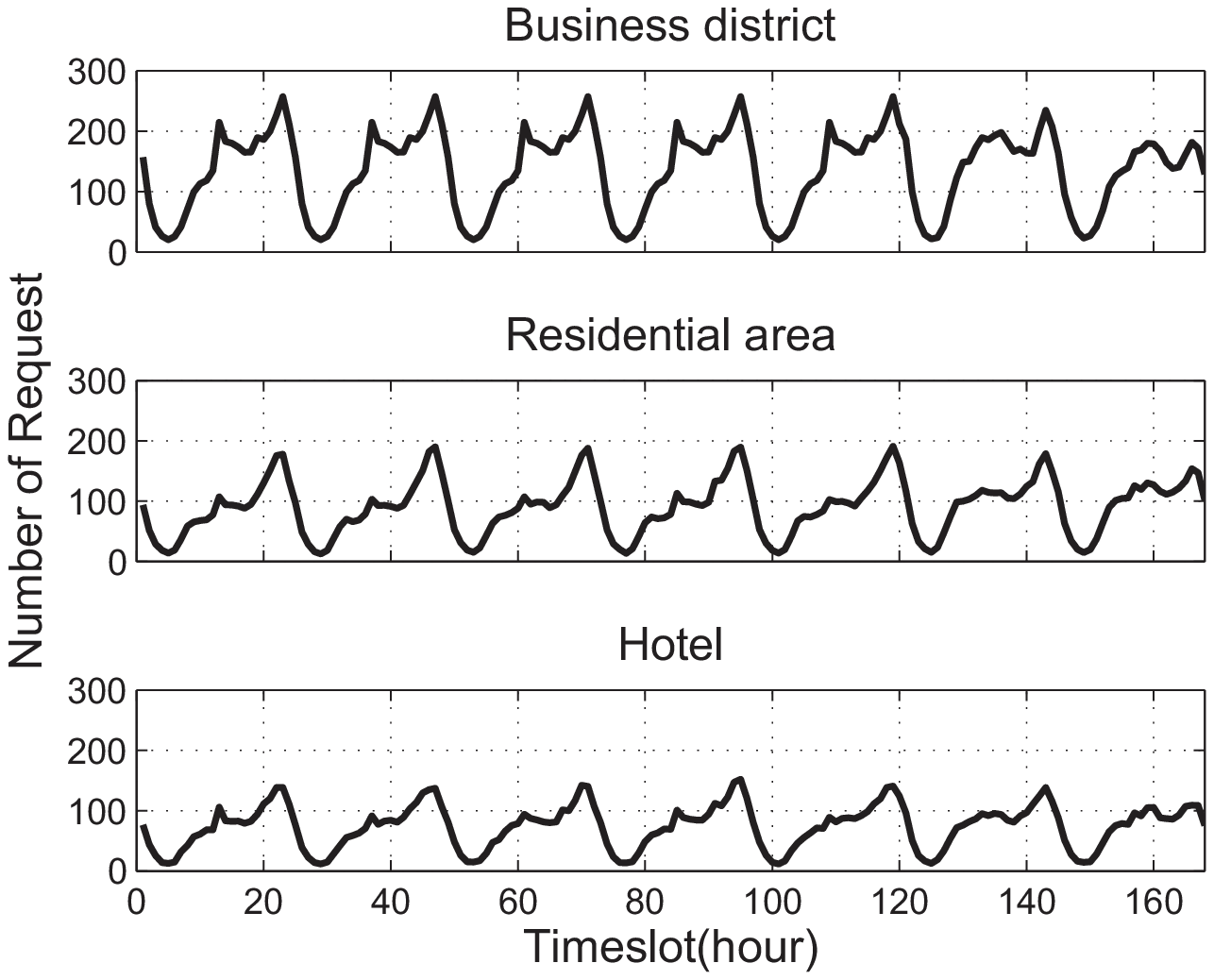}
          \caption{Number of mobile video requests in different locations over time.}
          \label{fig:request-over-time}
     \end{minipage}
\end{figure*}

We investigate the geographical distribution of requests.
 According to the longitude interval 0.01$^{\circ}$ and latitude interval 0.01$^{\circ}$ , Beijing will be divided into different locations. Every location can be abstracted as a 0.01$^{\circ}$ $\times$ 0.01$^{\circ}$ geographic location with an area of $0.72 km^2$. Each location has a PoI functionality label, which indicates the largest PoI functionality number of the location.
We count the number of requests issued in these locations. As shown in Fig.~\ref{fig:cdf-location-request}, we plot the CDFs of the number of requests issued in the locations at different times of a day, i.e., 6am--12am, 12am--6pm, and 6pm--12pm. Our observations are as follows: (1) More requests are issued at night than during the daytime, e.g., the number of requests from 6pm--12pm is $74\%$ greater than that from 12am--6pm. (2) A significant fraction of locations only have very few requests issued. These observations indicate that to serve mobile video requests, the edge network content delivery systems need to take the geographical request distribution into consideration, e.g., to allocate more resources to the locations with higher request density and proactively push content to the edge at the off-peak times.

\subsubsection{Multi-location Users in Different Locations}

We study the behaviors of multi-location users in different locations within one day, such as university, airport, railway station, scenery spot and business district. In Fig.~\ref{fig:jhmobiuser}, we plot the fraction of multi-location users over all video users recorded in our traces in these locations in one week.
Our observations are as follows: (1) These locations generally have a relatively stable multi-location user fraction of approximately $20\%$. (2) Some locations have lower multi-location user fractions than others, e.g., there are less users in university than at rail station.  (3) \emph{The fraction of multi-location users changes significantly over time in some locations}, e.g., the fraction in the business distinct drops from approximately $25\%$ on weekdays to $15\%$ on weekends. The reason is that the mobile video behaviors are highly correlated with the regular commute behaviors of users.

\subsubsection{Frequency Analysis of Periodical Request Patterns}

It is common for users to generate periodical requests, e.g., more video requests are issued at night. Such periodical request patterns can affect the edge network caching strategies, including content replication and resource allocation. As illustrated in Fig.~\ref{fig:request-over-time}, the curves represent the number of video requests issued in different locations in one week. The requests over time have different periodical patterns.

To specify the periodical request patterns, we use a frequency analysis approach \cite{wang2015understanding}, as follows: \begin{enumerate}

	\item Let $\mathbf{x} = (x_1,x_2,\ldots,x_N)^{T}$ denote the number of video requests over time, i.e., $x_i$ is the number of requests in time slot $i$. In our experiments, each time slot is $1$ hour, and we study the request samples in $1$ week, i.e., $N=168$.
	
	\item We perform DFT as follows,
	$$
		X[k] = \sum_{n=1}^{N}x_n e^{-2\pi i kn/N}, \label{eq:dft}
	$$
	where $X[k]$ is the frequency spectrum of sequence of requests $X$ in the time domain. A larger $X[k]$ indicates that the sequence has a stronger period of $k$.
	
	\item We study the amplitude of the frequency-domain sequence $X[k]$, in which amplitude and phase represent request volume and their peak-valley time, respectively.

\end{enumerate}

\begin{figure}[!t]
  \centering
    \includegraphics[width=.35\textwidth]{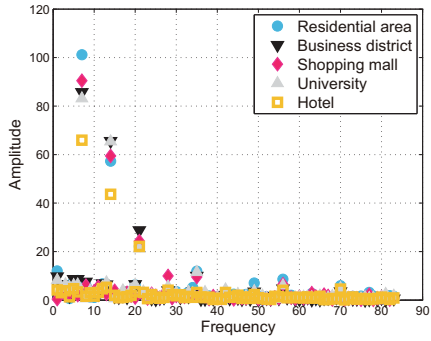}
  \caption{Frequency analysis of mobile video requests in different locations: amplitude versus frequency.}
  \label{fig:frequency}
\end{figure}

Fig.~\ref{fig:frequency} shows the discrete Fourier transform (DFT) results of the requests over time. In particular, we plot the amplitude versus the frequency of requests in different functionalities of locations. Our observations are as follows: (1) There are some major frequencies with large amplitudes, e.g., $k = 7, 14,$ and $21$, corresponding to $1$ day, $12$ hours and $8$ hours, respectively. This means that we can use the three frequency components to present the  time-domain traffic. Furthermore, we can leverage this property to predict the future traffic. (2) Different functionalities of locations also have different major frequency patterns. For example, the daily pattern is more obvious for the residential areas than the hotels, and the business areas have a strong period of $8$ hours. This observation indicates that \emph{the periodical patterns of mobile video requests are highly affected by the functional type of locations}, which can be utilized to distinguish locations with different functionalities.

\subsection{Analysis on Content Video: An Entropy Approach}
\label{sec:entropy1}

We study how different videos are actually requested in different locations. To this end, we use an entropy analysis approach. Motivated by the entropy calculation in information theory \cite{brodersen2012youtube,li2014geographic,Shannon1948A}, we define a video request entropy and a location request entropy.

\subsubsection{Geographical Video Request Entropy}

The geographical video request entropy $H^V(v)$ is defined as follows:
$$
	H^V(v) = - \sum_{l \in \Region_v} \frac{n_{vl}}{\sum_{j\in \Region_v} n_{vj}} \log \frac{n_{vl}}{\sum_{j\in \Region_v} n_{vj}}, \label{eq:video:entropy}
$$
where $H^V(v)$ is the geographical video request entropy for video $v$, $\Region_v$ is the set of locations (e.g., locations defined previously) where video $v$ has been requested, and $n_{vj}$ is the number of requests for video $v$ in location $j$.  A lower value of video request entropy indicates that the video's requests are more diversely distributed across different locations, thereby affecting the caching strategies.

\subsubsection{Location Request Entropy}

The next entropy is location request entropy, which reflects the diversity of videos requested in a particular location. The location request entropy is calculated as follows:
$$
	H^L(l) = -\sum_{v \in \Videos_l} \frac{n_{vl}}{\sum_{j \in \Videos_l}n_{jl}} \log \frac{n_{vl}}{\sum_{j \in \Videos_l}n_{jl}}, \label{eq:location:entropy}
$$
where $\Videos_l$ is the set of videos requested in location $l$. A larger location request entropy value indicates that the videos requested in the location are more diverse. For the caching strategy, a location with a larger location request entropy generally requires more content items to be replicated to serve the users. 

We can compare geographical video request entropy (location request entropy) to evaluate their request patterns given fixed total videos and locations. However, it is unfair to compare geographical video request entropy (location request entropy) directly if the total number of locations where video requests are issued is different (each location has different unique videos). Once any additional location is involved, the entropy will be increased \cite{chen2014fake}. For example, the requested videos with more locations tend to have larger geographical video request entropy than videos with fewer requesting locations. To overcome this ambiguity, the two entropies have been normalized in our measurement.

\subsubsection{Entropy Analysis}
\label{sec:entropy}
In this section, we will conduct entropy analysis from two perspectives: video and location.

\begin{figure}[!t]
	\centering
		\subfigure[Non-normalized]{
			\label{fig:entropy-geo:a}
			\includegraphics[width=0.47\linewidth]{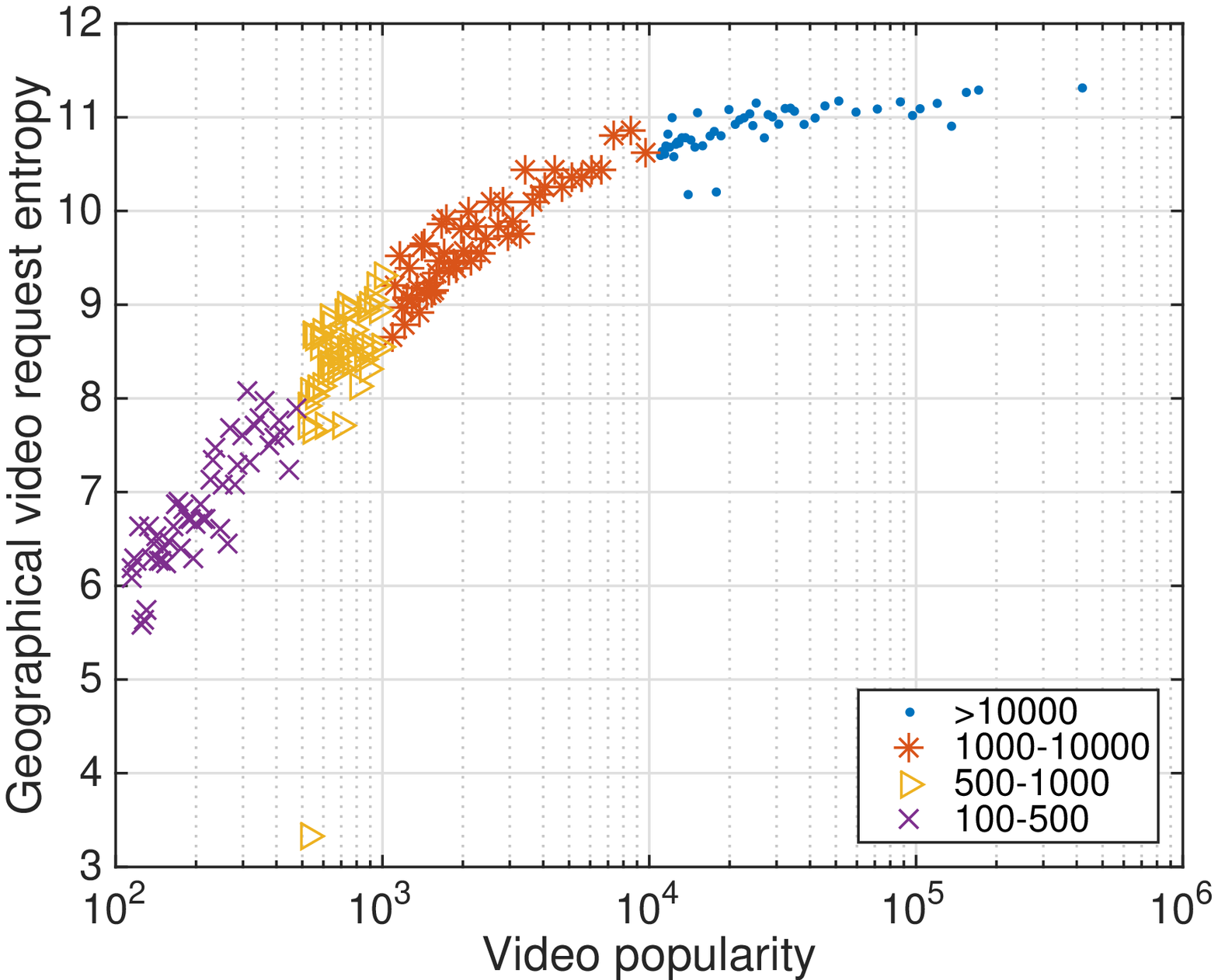}}
		\hfill
		\subfigure[Normalized]{
			\label{fig:entropy-geo:b}
			\includegraphics[width=0.47\linewidth]{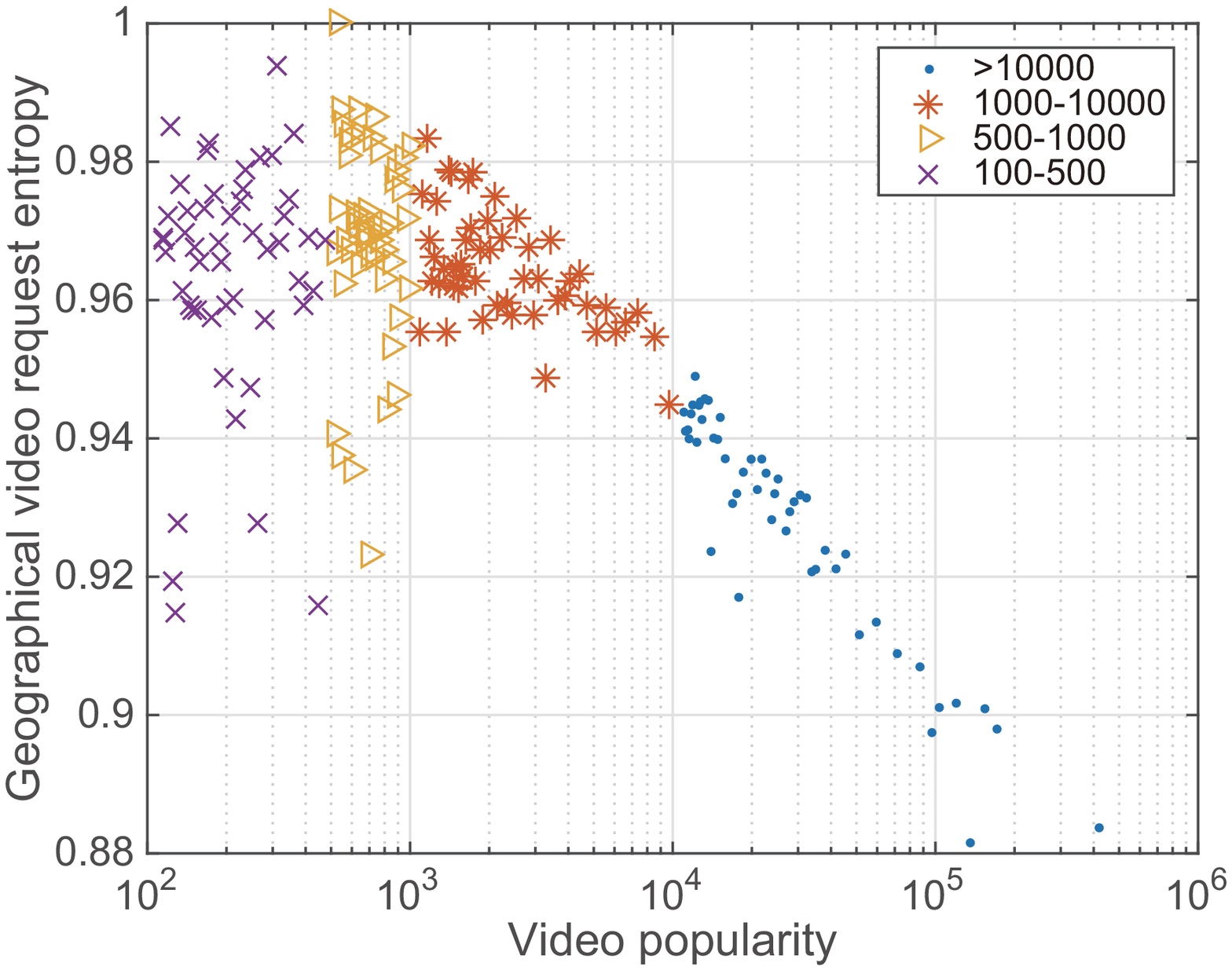}}
	\caption{Geographical video request entropy versus video popularity.}
	\label{figure{fig:entropy-geo}}
\end{figure}

From the video perspective, we primarily use geographical video request entropy. We first divide videos into four grades according to the requested times (i.e., video popularity) during two weeks, and then we select $50$ videos from each grade randomly and compute the geographical video request entropy for these videos. Fig.~\ref{fig:entropy-geo:a} shows that the geographical video request entropy increases as the video popularity increases. This result is consistent with the general understanding. The more popular the video is, the larger its geographical video request entropy is. Consequently, popular videos receive requests from almost everywhere (global distributions), whereas unpopular videos only receive requests from some specific locations (local distributions). Fig.~\ref{fig:entropy-geo:b} shows the normalized geographical video request entropy. Interestingly, this figure shows a different result that the more popular the video is, the smaller the normalized entropy is. The reason is that \emph{although the requests for a popular video are requested from more locations, the request distribution of these locations is more skewed compared with that of unpopular videos}.

\begin{figure}[htbp]
    \centering
    \includegraphics[width=0.35\textwidth]{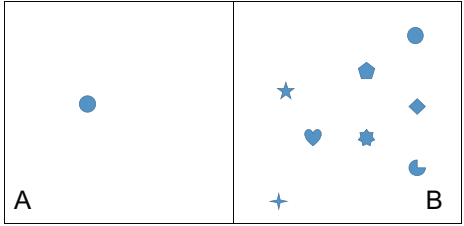}
     \caption{Location request entropy schemetic diagram.}
     \label{fig:entropy-sche}
\end{figure}

\begin{figure*}[!t]
	\centering
		\subfigure[Location request entropy distribution]{
			\label{fig:entropy-dist}
			\includegraphics[width=0.22\linewidth]{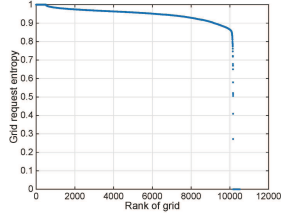}}
			\hfill
		\subfigure[Location request entropy versus number of PoI labels]{
			\label{fig:entropy-poi}
			\includegraphics[width=0.22\linewidth]{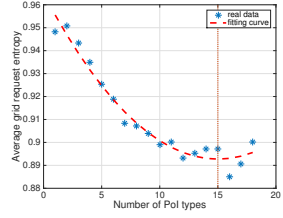}}
			\hfill
		\subfigure[Location request entropy versus user mobility intensity]{
			\label{fig:entropy:mobility}
			\includegraphics[width=0.22\linewidth]{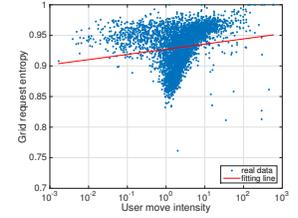}}
	\caption{Content request geographical distribution: entropy analysis.}
	\label{fig:entropy}
\end{figure*}

From the location perspective, we primarily use the normalized location request entropy. We study the distribution of the location request entropies. In Fig.~\ref{fig:entropy-dist}, we plot the location request entropies $H^L(l)$ of $10514$ locations versus the rank of the locations. The results are calculated based on our 2-week traces. We observe that the normalized location request entropy distribution is almost a straight line without the smallest locations, ranging from $0.8$ to $1$. To better understand these values, Fig.~\ref{fig:entropy-sche} shows the corresponding schematic diagram, where different shapes represent different unique videos. Location $A$ only requests one video each time; thus, the entropy is $0$. However, users in location $B$ issue eight requests for eight different videos and the distribution is uniform; thus, the corresponding request entropy is $1$. 
Therefore, the fewer video requests and strongly skewed distributions result in the smallest entropy of the locations. Intuitively, locations with more unique videos have smaller entropy, which have more skewed distributions.
Considering cache strategies, \emph{LFU is better for locations with smaller location entropy since there are many different requested videos at each time, whereas LRU is better for locations with larger location entropy}.

We next investigate whether the normalized location request entropy is affected by the characteristics of the location. In particular, we study the correlation between the location request entropy and the number of PoI functionality labels of a location, e.g., residential area. As shown in Fig.~\ref{fig:entropy-poi}, each sample is the average normalized entropy of locations versus the number of PoI labels of these locations. Our observations are as follows: (1) Locations with a larger number of PoI labels typically have smaller location request entropies, indicating that a location with more ``functionalities'' has a more skewed request distribution accompanied by more diverse video requests. (2) The relationship approximately follows the quadratic function $y={0.0003}x^2-0.0096x+0.9648$, indicating that the location with a particular number (nearly fifteen) of PoI labels has the smallest entropy.


Finally, we investigate the impact of user mobility on the normalized location request entropy. We define the user mobility intensity of a location as the mean of all multi-location users who have requested videos in that location. In this experiment, locations with no user movement are not considered. In Fig.~\ref{fig:entropy:mobility}, each sample is the location request entropy versus the user mobility intensity. As shown in this figure, as the user mobility intensity increases from $10^{0}$ to $10^{3}$, the normalized location request entropy gradually increases. We fit the samples into the function $y= 0.0085\log(x)+0.9273$, implying that user mobility is also a factor for the content diversity. In contrast to user number, the user mobility intensity has a positive impact on location request entropy. \emph{The larger the user mobility intensity is, the larger the location request entropy is and the less the unique video number of the location is.} It is inferred that multi-location users are more likely to request popular videos without increasing the unique video number. One possible reason is that the time of the multi-location users is fragmented such that they are more interested in popular videos. Thus, LFU is more suitable for multi-location users. We will verify the results in Sec.~\ref{sec:edgenet}.

\section{Mobile Video Requests Affected by User Mobility Behaviors} \label{sec:userpatterns}

In this section, we study what drives the previous request patterns. Particularly, we focus on mobile video user behaviors. In the following experiments on multi-location users, our results are the average results of fourteen days.

\subsection{Mobility Intensity Analysis}

In our experiments, we only study the behaviors of \emph{active users} who requested at least ten videos daily in our 2-week traces. Among these $9,576$ \emph{active users}, we have $30\%$ \emph{multi-location users} and $70\%$ \emph{simple-location users}, which are defined previously.

\subsubsection{Movements and Locations Visited}

We first study the mobility intensity of the multi-location users. In Fig.~\ref{figure{fig:locations:a}}, we plot the fraction of users versus the number of ``movements'', i.e., the number of requests issued in different locations in \emph{one day}. We observe that the number of movements is generally in the range $[1,30]$, and the range $[2,3]$ has the largest fraction of users. The results are quite similar for weekdays and weekends. We next study the number of locations where the requests are issued. In Fig.~\ref{figure{fig:locations:b}}, the bars are the fraction of users versus the number of locations where videos are requested in one day. As shown in this figure, as many as $50\%$ of the multi-location users only issued video requests at $2$ locations, and $80\%$ of the users only requested videos from less than $4$ locations. These results indicate that \emph{it is common for multi-location users to request videos from different locations, but the number of locations (per user) is quite limited}. It provides some basic characteristics to capture the trajectory of multi-location users.

\begin{figure}[!t]
	\centering
		\subfigure[Movement number]{
			\label{figure{fig:locations:a}}
			\includegraphics[width=0.47\linewidth]{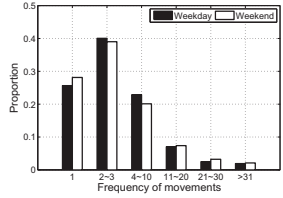}}
		\hfill
		\subfigure[Location number]{
			\label{figure{fig:locations:b}}
			\includegraphics[width=0.47\linewidth]{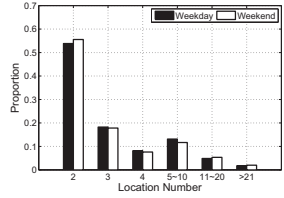}}
	\caption{Statistics of movement of mobile users.}
	\label{figure{fig:locations1}}
\end{figure}

\subsubsection{Distance and Interval of Movements}

\begin{figure}[!t]
	\centering
		\subfigure[Distance]{
			\label{fig:distance}
			\includegraphics[width=0.47\linewidth]{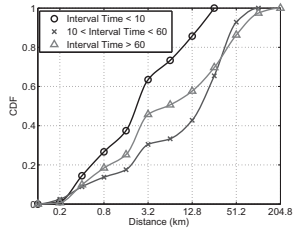}}
		\hfill
		\subfigure[Interval]{
			\label{fig:interval}
			\includegraphics[width=0.47\linewidth]{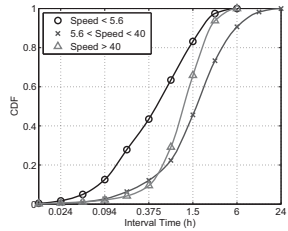}}
	\caption{CDFs of distances and intervals of consecutive requests.}
	\label{figure{fig:locations}}
\end{figure}

We further measure the cumulative distribution of the distances between consecutively visited locations with different time intervals. Fig.~\ref{fig:distance} plots the CDFs of distances between locations where users consecutively request mobile videos. In detail, we select $3$ intervals to divide users into the same order: $[0,10)$ min, $[10,60)$ min, and $[60,\infty]$ min. We observe that when the interval is  shorter than $10$ min, the distance is much shorter than that with the other intervals. However, as the interval time increases, the distance does not always become longer. The small time interval indicates that users frequently move between different locations. It is inferred that most users move between $2$ or $3$ locations in a small time interval. 

We also study the intervals between consecutive mobile video requests. In Fig.~\ref{fig:interval}, we plot the interval between consecutive requests of users with different movement speeds. We choose two reference speeds: the average speed of walking (i.e., $5.6$ km/h) and the average speed of subway (i.e., $40$ km/h). As shown in this figure, when the speed is less than $5.6$ km/h, most of the request intervals are small. For example, $80\%$ of the request intervals are issued within $1.5$ hours, whereas only $40\%$ of the request intervals are issued in $1.5$ hours for the movement speed of $[5.6,40)$ km/h. These observations indicate that the mobility speed of users also affects the request patterns. 
Moreover these results largely depend on vehicles, which determine the enroute time.

\subsection{Migration Patterns}

For the multi-location users who request videos in different locations, we study their migration patterns, i.e., how they move across different locations. 

\subsubsection{Location Migration Pattern}

\begin{figure}[!t]
	\centering
		\includegraphics[width=.5\linewidth,angle = 90]{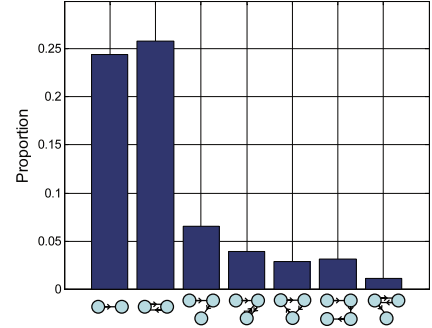}
	\caption{Fraction of migration patterns.}
	\label{fig:mig-pattern}
\end{figure}

According to our previous observations, users only request mobile videos in a small number of locations. We study how they move across these locations. In Fig.~\ref{fig:mig-pattern}, we plot the fraction of users who share the same migration patterns across different locations. In this figure, we plot the most popular $7$ migration patterns, which contribute $70\%$ of all the migrations between locations. We observe that \emph{moving between two particular locations constitutes almost $50\%$ of the migrations}. Additionally, there are migration patterns across $3$ and $4$ locations. These results provides us with the basic characteristics to construct connections between different locations for achieving caching cooperation strategies.

\subsubsection{Location Type Migration Pattern}

We study the migration between different functionalities of locations. Based on the PoI information used in our previous measurement studies, we calculate the number of migrations of users from one functionality of location to another functionality of location. As summarized in Table \ref{tab:migration}, each entry is the number of migrations in two weeks, e.g., there are $2,223$ migrations from the hospital areas to the business areas. We observe that (1) it is common for users to move between locations with the same PoI type, e.g., business to business, and (2) there are large migration numbers between some specific pairs of location functionalities, e.g., the largest migration number occurs between residential areas and business areas.

\begin{table}[!t]
\setlength{\abovecaptionskip}{5pt} 
\setlength{\belowcaptionskip}{5pt}
	\tiny
	\begin{center}
	\caption{\label{tab:migration}Migration Matrix.}
	\begin{tabular}{p{.7cm} p{.7cm} p{.7cm} p{.7cm} p{.7cm} p{.7cm} p{.7cm} p{.7cm}}
	\toprule
		{From / to} & Business & Hospital & Resident & Campus & Scenery & Shopping & Hotel\\
	\midrule
		Business & 4908 & 2205 & 5114 & 1379 & 595 & 1082 & 657\\
		Hospital & 2223 & 1741 & 3479 & 802 & 394 & 698 & 360 \\
		Resident & 5145 & 3425 & 9994 & 1787 & 995 & 1727 & 907 \\
		Campus & 1369 & 797 & 1743 & 843 & 230 & 367 & 222 \\
		Scenery & 596 & 399 & 984 & 215 & 183 & 187 & 123 \\
		Shopping & 1101 & 692 & 1671 & 358 & 234 & 494 & 169 \\
		Hotel & 616 & 367 & 928 & 214 & 114 & 202 & 213 \\
	\bottomrule
	\end{tabular}
	\end{center}
\end{table}

\section{Edge Network Content Delivery for Mobile Video Streaming} \label{sec:edgenet}

In this section, we compare the effectiveness of Wi-Fi APs and cellular base stations-based edge content delivery solutions, and we discuss the potential improvement to today's wireless networks to enhance mobile video streaming. We first study whether the request intensity in different locations matches the number of edge network infrastructures; we then present the difference between cellular and Wi-Fi on spatial and temporal patterns. In particular, we focus on the effects on caching performance of influencing factors, including different strategies, request density, video and user diversity and user mobility.

\subsection{Request Coverage by Edge Network Infrastructure}
\label{infrastructure}

To answer the question of whether today's edge network infrastructure can appropriately satisfy the mobile video streaming demand, we measure the distance between users and their nearest infrastructure, and we compare the differences of distribution between requests and edge network infrastructure.

\subsubsection{Distance between Requests and APs/BSes}

\begin{figure}[!t]
	\centering
		\subfigure[CDF of distances between requests and nearest AP/BS]{
			\label{fig:req-dist}
			\includegraphics[width=0.47\linewidth]{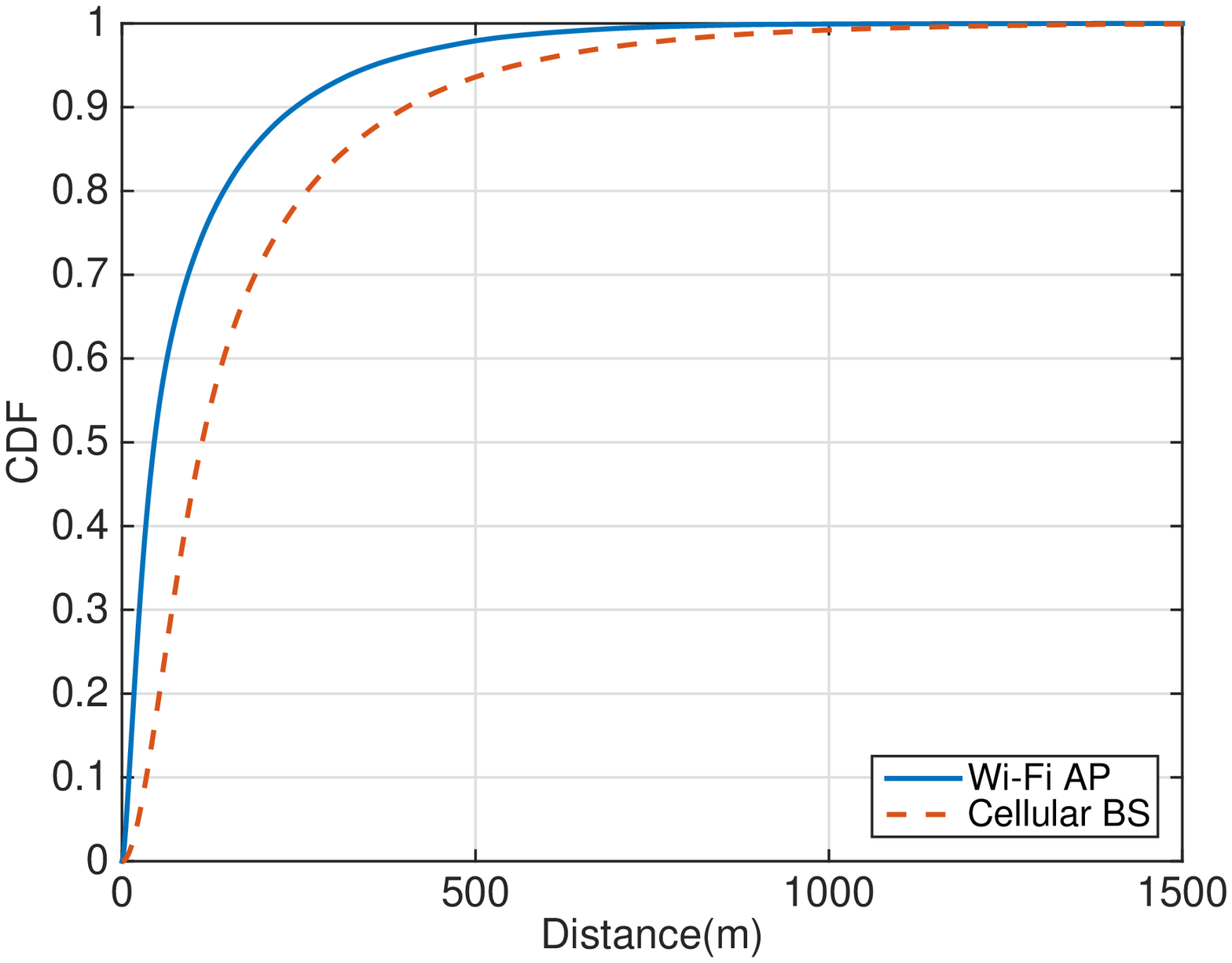}}
		\hfill
		\subfigure[CDF of distance differences between request and Wi-Fi APs and cellular BSes]{
			\label{fig:distance_gap}
			\includegraphics[width=0.47\linewidth]{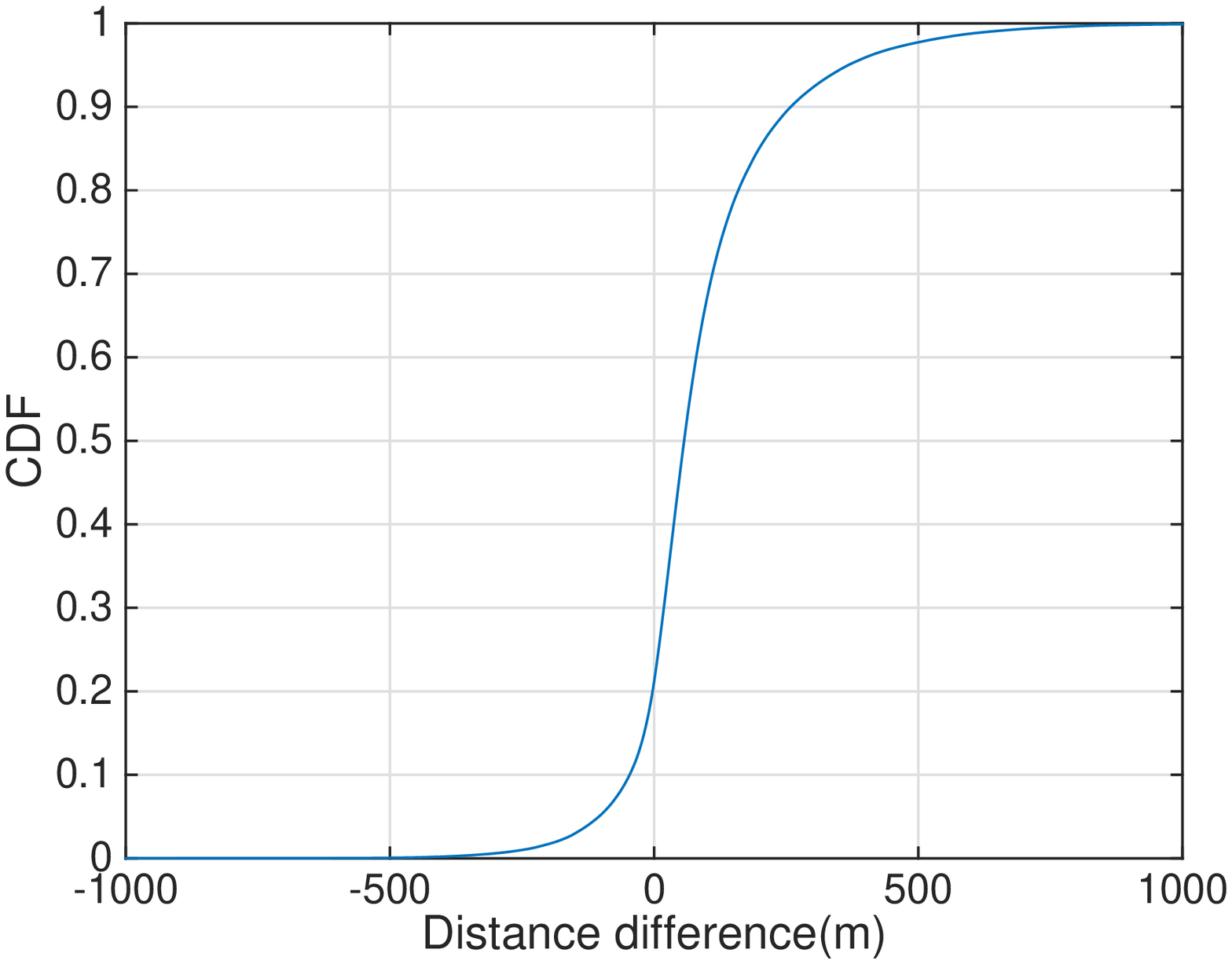}}
	\caption{Request coverage by edge-network infrastructure.}
	\label{figure{fig:locations2}}
\end{figure}

We investigate how the mobile video requests can be served by nearby edge network infrastructures, including the Wi-Fi APs (i.e., the smartrouter mode) and cellular base stations (i.e., the femtocell mode). In particular, we study how far away users can find an AP or base station to download videos. Fig.~\ref{fig:req-dist} plots the CDFs of the distances between the requests and the nearest Wi-Fi APs or cellular BSes that can potentially serve them. We observe that over $95\%$ of the mobile video requests can at least find a Wi-Fi AP within $500$ meters or a cellular BS within $750$ meters, indicating that edge network video content delivery is promising. 
We further compare the distance between a video request and the nearest Wi-Fi AP, and the distance between the same request and the nearest cellular BS. In Fig.~\ref{fig:distance_gap}, a distance gap larger than $0$ suggests that the distance for cellular base station is larger than the distance for Wi-Fi AP. We observe that over $80\%$ of the distance differences are larger than $0$, suggesting that Wi-Fi APs are generally closer to users.

\subsubsection{Request Intensity versus Number of APs/BSes}

\begin{figure}[!t]
	\centering
		\subfigure[Selected area]{
			\label{fig:map}
			\includegraphics[width=0.47\linewidth]{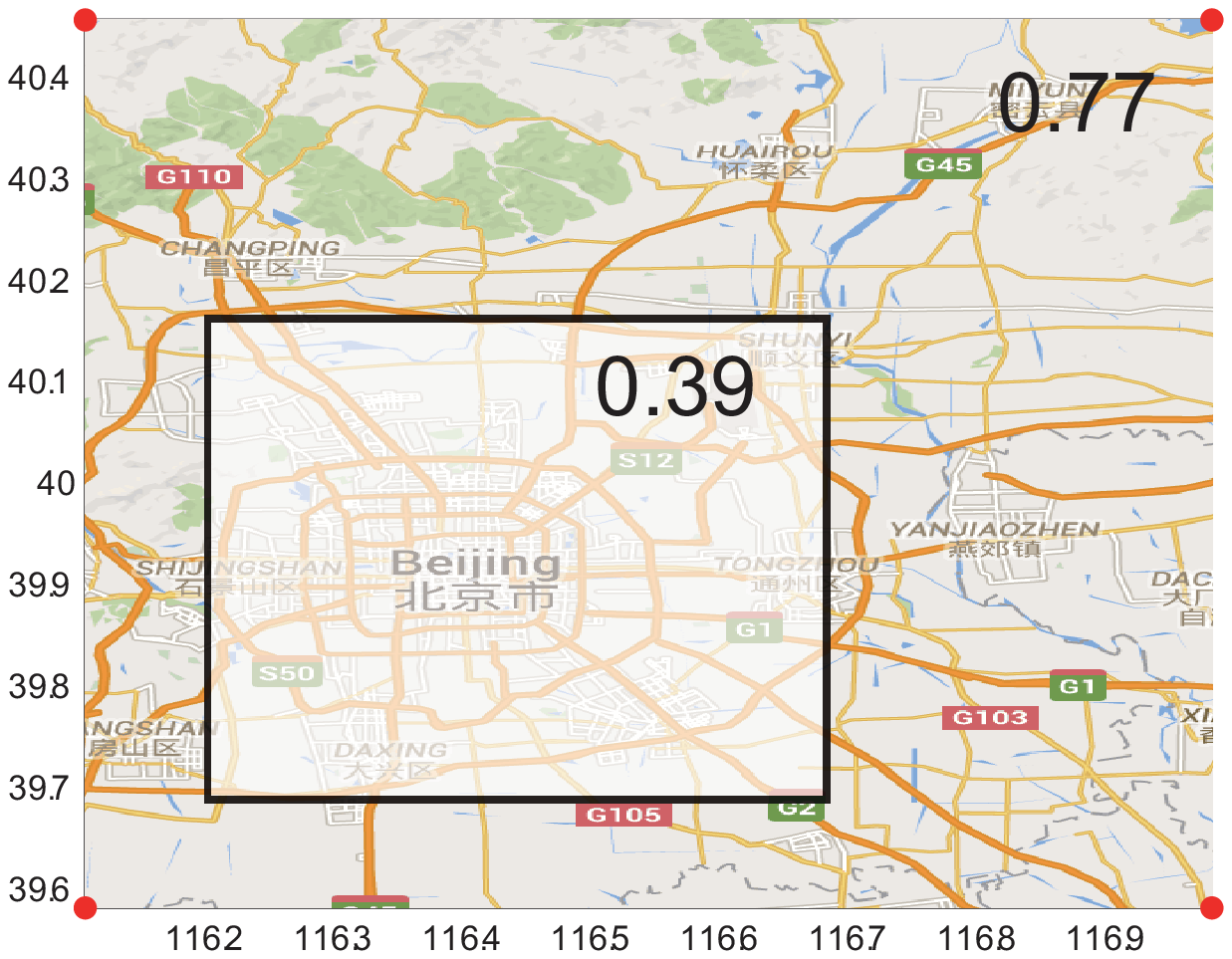}}
		\hfill
		\subfigure[Scatter distribution]{
			\label{fig:request}
			\includegraphics[width=0.47\linewidth]{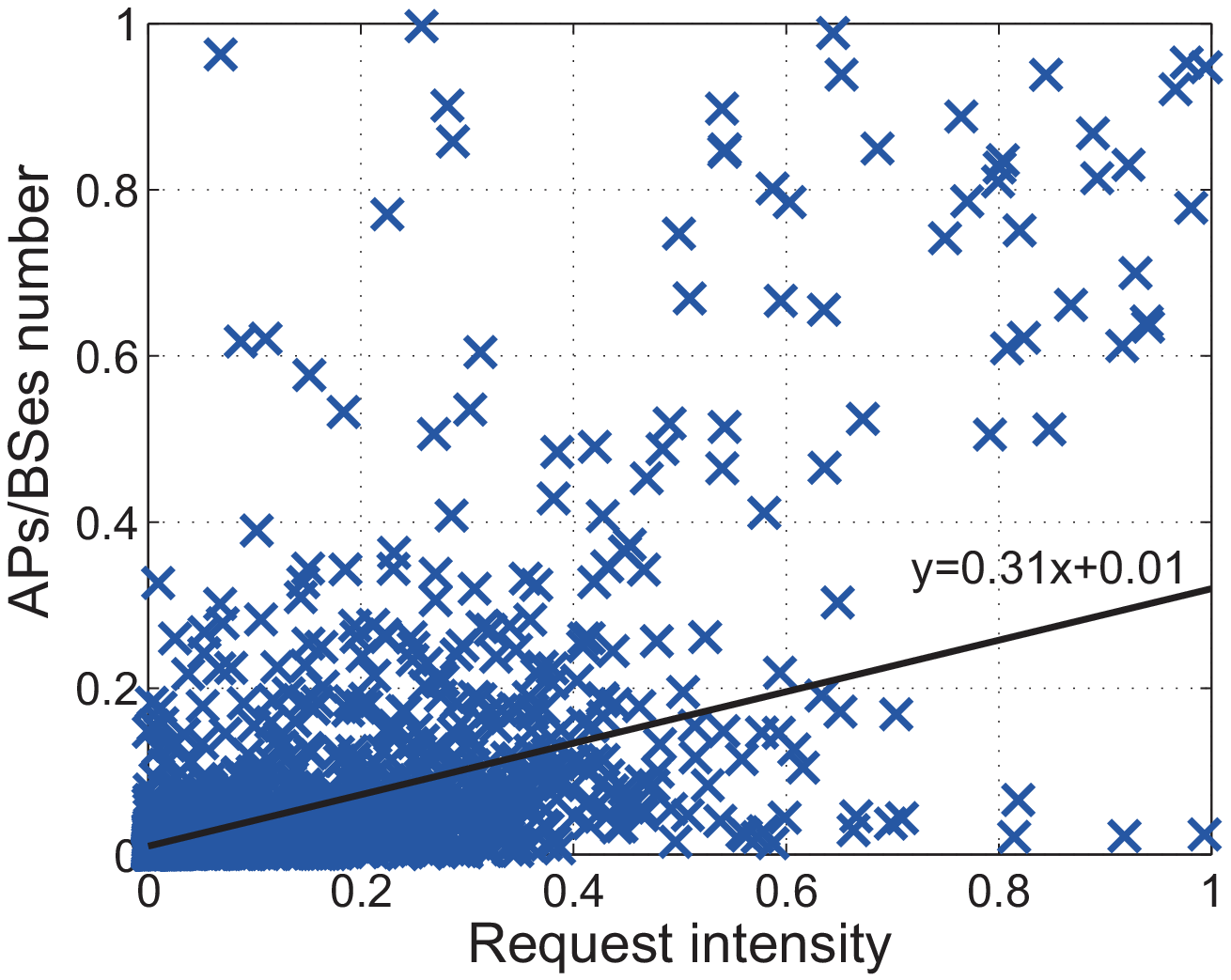}}
	\caption{Request intensity versus number of APs/BSes.}
	\label{fig:request_distribution}
\end{figure}

We also investigate the request intensity (number of requests in different locations) and the number of Wi-Fi APs and cellular base stations in the entire city under two assumptions: all the requests have the same cost, and all APs/BSes have the same power.
We use the max-min method to normalize request intensity and number of APs/BSes ranging from 0 and 1. In particular, we investigate whether the requests and the APs/BSes share the same distribution, e.g., there are more APs/BSes if there are more requests in the same location. To this end, we calculate the cosine similarity between the two, i.e., $\Requests \cdot \Infra$, where $\Requests$ is the normalized vector of the numbers of requests in the locations and $\Infra$ is the normalized vector of the numbers of the APs or BSes at the same locations. A large similarity indicates that the request intensity matches the number of APs/BSes.
We observe that the similarity is higher than $0.77$, which is considered to indicate a significant similarity. Interestingly, only considering the centralized 30\% area that occupies more than 80\% of the total requests, the similarity is less than 0.39, as shown in Fig.~\ref{fig:map}. Fig.~\ref{fig:request} shows the comparison between them. These results imply that a marked difference exists between request intensity and number of APs/BSes, particularly in high-intensity locations where APs/BSes are generally unable to satisfy the requests. It suggests that video service providers should deploy more APs/BSes to better satisfy the users' quality of experience.

\subsubsection{Wi-Fi/Cellular Stability Analysis}
\begin{figure}[!t]
	\centering
		\subfigure[CDF of KL distance]{
			\label{fig:KL_distance}
			\includegraphics[width=0.47\linewidth]{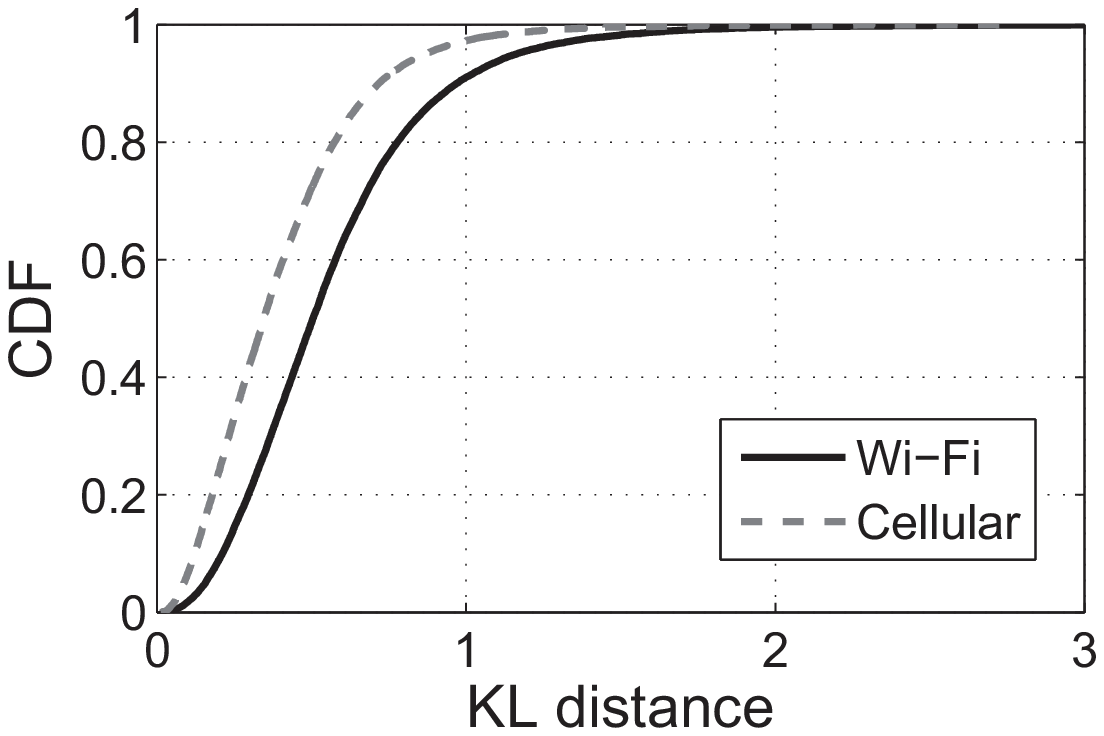}}
		\hfill
		\subfigure[Example]{
			\label{fig:KL_example}
			\includegraphics[width=0.47\linewidth]{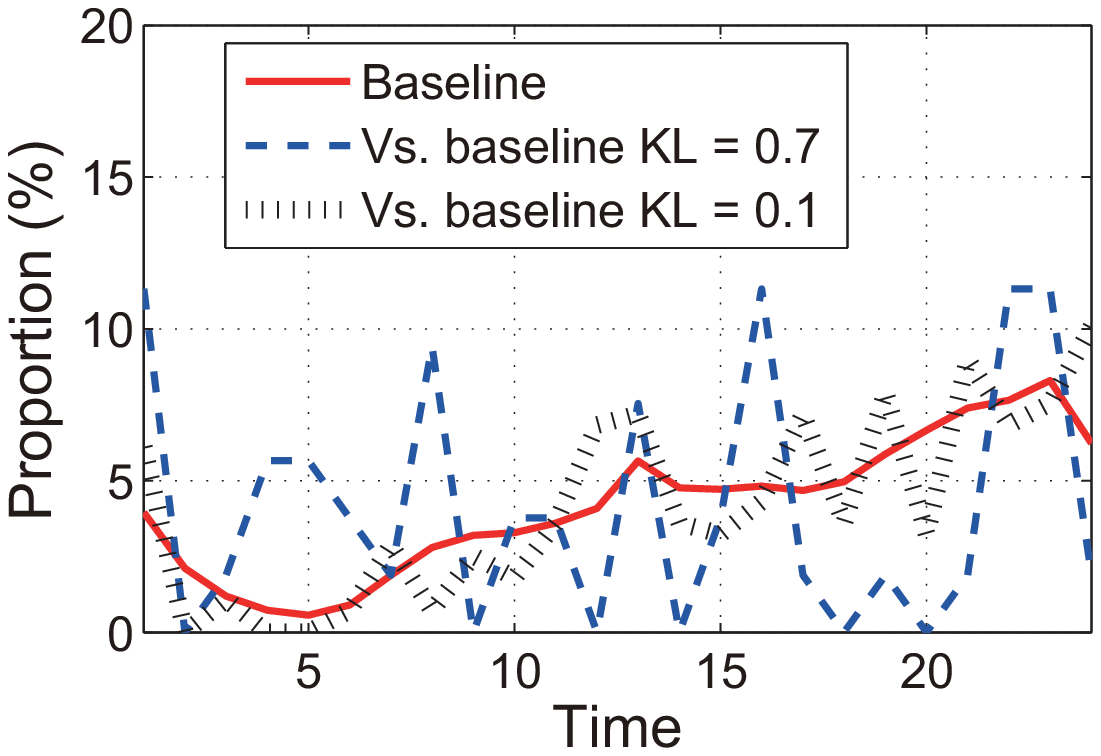}}
	\caption{KL distance between daily request distributions.}
	\label{fig:KL}
\end{figure}

From each Wi-Fi AP and cellular base station perspective, we are interested in the following question. Does the request time distribution of Wi-Fi/cellular follow the global request distribution? To answer this question, we measure the divergence between the time distribution of global requests and single Wi-Fi/cellular requests. To this end, we use the Kullback-Leibler (KL) distance to measure the distance between two distributions, which is defined on two distributions $P$ and $Q$ as follows:
$$
{D_{KL}}(P\parallel Q) = \sum\limits_{t \in T} {P(t)\log \frac{{P(t)}}{{Q(t)}}},
$$
where $T$ refers to the set of time, $P$ is the Wi-Fi/cellular distribution of request time on a particular day, and $Q$ is the distribution of global request. The KL distance is a non-negative value. It represents the number of extra bits necessary to encode samples from $P$ when using a code based on $Q$, rather than directly based on $P$. The smaller the value is, the closer the two distributions are. Fig.~\ref{fig:KL_distance} depicts the CDF of KL distance over Wi-Fi and cellular, which conveys that the daily request distributions of cellular relatively follow the global request distribution. The reason is that Wi-Fi APs attract users mostly from a smaller location where users have particular interests in the content. 
Fig.~\ref{fig:KL_example} illustrates two instances of KL distance. It shows that when KL distance equals $0.1$, the distribution is similar to the baseline.

\subsection{Performance of Content Caching by Edge Networks}
\label{caching_evaluation}

In this section, we evaluate the quality of user experience in current mobile video systems. We assume that a user has a better quality of experience (lower delay) when he is served by the edge cache of a Wi-Fi AP or cellular base station.
Thus, we build a discrete trace-driven simulator to evaluate the cache hit rates of conventional caching strategies for Wi-Fi APs and cellular BSes. We simulate mobile video requests, following the records in the real-world traces. We also use the positions of the APs and BSes recorded in our traces to simulate the edge network infrastructure. 

\subsubsection{Experimental Setup}

In the simulation experiments, we assume that the average video size $S$ is unit \cite{golrezaei2012femtocaching,zhou2015video,hachem2015content} and the evaluation criterion is the total cache hit rate. We use the $2$-week records of users' requests to drive the simulation, and we let the requests be served by the nearest Wi-Fi APs or cellular BSes. All the APs/BSes have the same cache capacity $C$ and the default cache capacity is $20$ (items). We set the concurrency of APs/BSes as $20/100$ and the bandwidth as $20S/100S$ for APs/BSes to limit the max transmission number per unit time \cite{tripathi2016optimizing}. The radius of each AP/BS is $100m/500m$ \cite{herlich2014optimal}. The parameters of the experiments are summarized in Table~\ref{tab:setup}.

In our experiments, we use the following conventional caching strategies: (1) Least recently used (LRU). It discards the least recently used content item first when the cache is full. (2) Least frequently used (LFU). It discards the least frequently used item first when the cache is full. (3) Random replacement (RR). It randomly selects a candidate item and discards it when necessary.

\begin{table}[!t]
\setlength{\abovecaptionskip}{5pt} 
\setlength{\belowcaptionskip}{5pt}
\small
\begin{center}
	\caption{Simulation Parameters Setting.}\label{tab:setup}
	\begin{tabular}{lp{0.2\textwidth}}
	\toprule
		\multicolumn{1}{c}{Parameter} & \multicolumn{1}{c}{Value} \\
	\midrule
		\multicolumn{1}{l}{Average video size $S$} & \multicolumn{1}{c}{Unit} \\
		\multicolumn{1}{l}{Cache capacity $C$} & \multicolumn{1}{c}{$20$ (default)} \\
		\multicolumn{1}{l}{Concurrency of APs/BSes} & \multicolumn{1}{c}{$20/100$} \\
		\multicolumn{1}{l}{Bandwidth of APs/BSes} & \multicolumn{1}{c}{$20S/100S$} \\
		\multicolumn{1}{l}{Radius of APs/BSes} & \multicolumn{1}{c}{$100m/500m$} \\
	\bottomrule
	\end{tabular}
\end{center}
\end{table}

\subsubsection{Hit Rate versus Capacity}

\begin{figure}[!t]
	\centering
		\includegraphics[width=.7\linewidth]{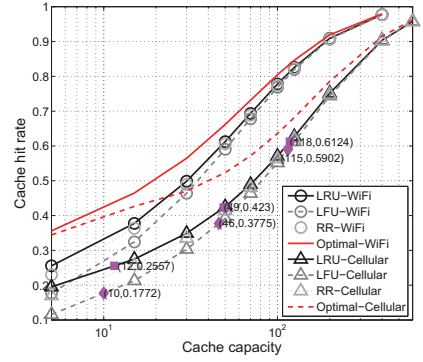}
	\caption{Cache hit rate under different cache capacity.}
	\label{fig:cache}
\end{figure}

We first study the impact of the cache capacity on the cache hit rate. Fig.~\ref{fig:cache} shows the cache hit rates of different caching strategies for both Wi-Fi and cellular networks by varying the cache capacity from $5$ to $600$. Our observations are as follows: (1) The cache hit rates in Wi-Fi APs are generally larger than that in cellular BSes, e.g., to reach the same cache hit rate of $0.25$ ($0.42$, $0.61$) with LRU, the average cache capacity of the Wi-Fi APs is $5$ ($20$, $50$), whereas it is $12$ ($49$, $118$) for cellular BSes. Additionally, the result is similar to LFU. (2) LRU, LFU and RR achieve similar cache hit rates, particularly when the cache capacity is large. As the cache capacity increases, the probability of a new item being discarded in RR gradually decreases, resulting in similar performance with LRU. Since the cache contains increasingly more items, all of the strategies achieve high cache hit rates.

\subsubsection{Impact of Request Density}

\begin{figure}[!t]
	\centering
		\subfigure[Cellular]{
			\label{fig:request_BS}
			\includegraphics[width=0.47\linewidth]{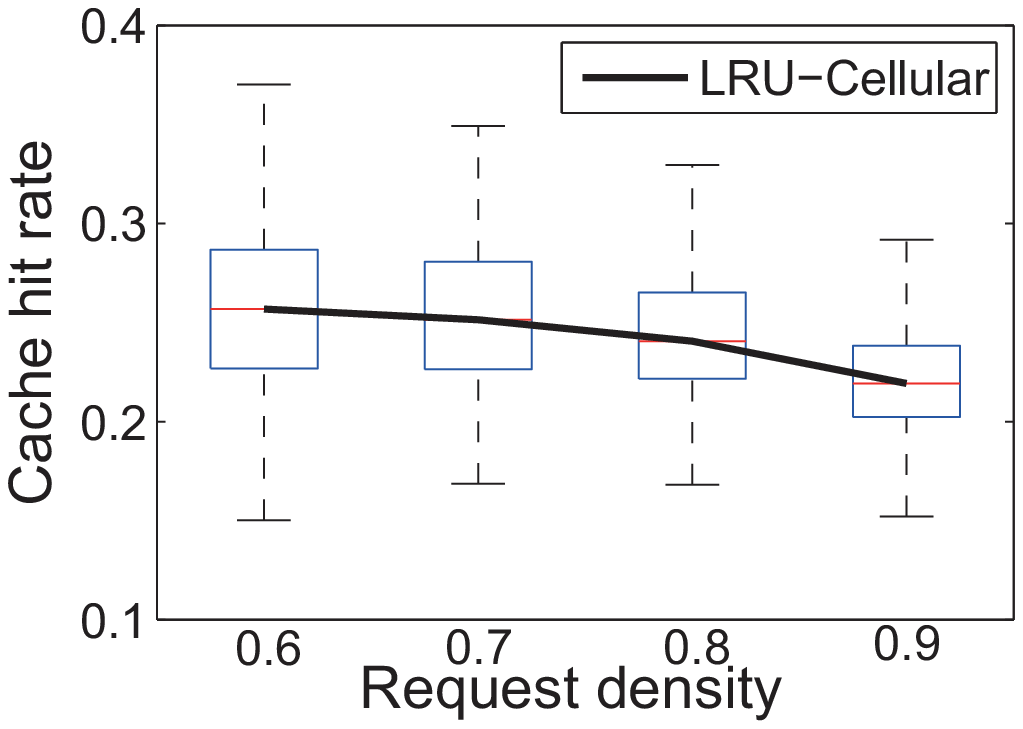}}
		\hfill
		\subfigure[Wi-Fi]{
			\label{fig:request_wifi}
			\includegraphics[width=0.47\linewidth]{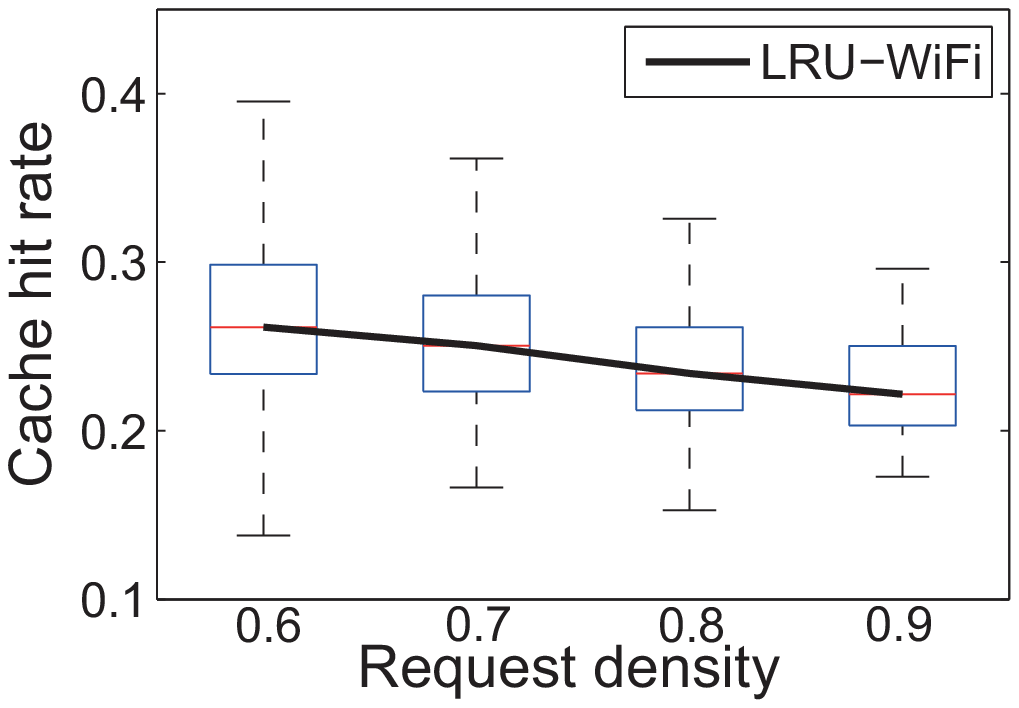}}
	\caption{Cache hit rate versus request density.}
	\label{fig:request_intensity}
\end{figure}

According to previous measurement studies, different locations present different levels of requests. We study the caching performance for locations with different request density levels. Fig.~\ref{fig:request_intensity} shows the box-plots of cache hit rates of (a) APs and (b) BSes with different normalized request levels--the request density level is normalized in $[0,1]$. We use the LRU strategy with a capacity of $20$ unless noted otherwise. We observe only a slight decrease in the cache hit rate with increasing request density, and the variation becomes smaller when there are more requests. These results suggest that the caching strategies are relatively insensitive to the request density.

\subsubsection{Video and User Diversity}

\begin{figure}[!t]
	\centering
		\subfigure[Cellular caching]{
			\label{fig:diversity_BS}
			\includegraphics[width=0.47\linewidth]{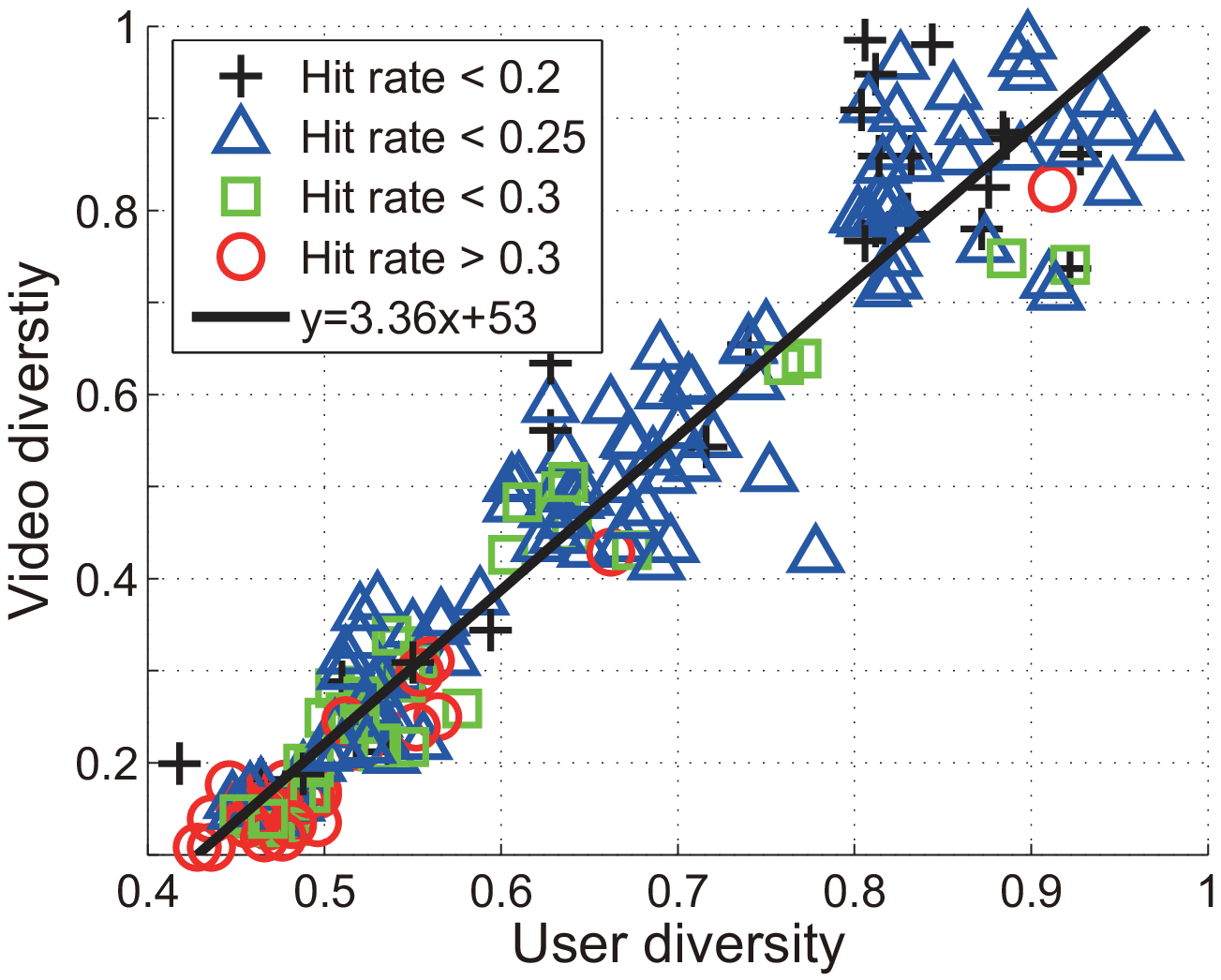}}
		\hfill
		\subfigure[Wi-Fi caching]{
			\label{fig:diversity_wifi}
			\includegraphics[width=0.47\linewidth]{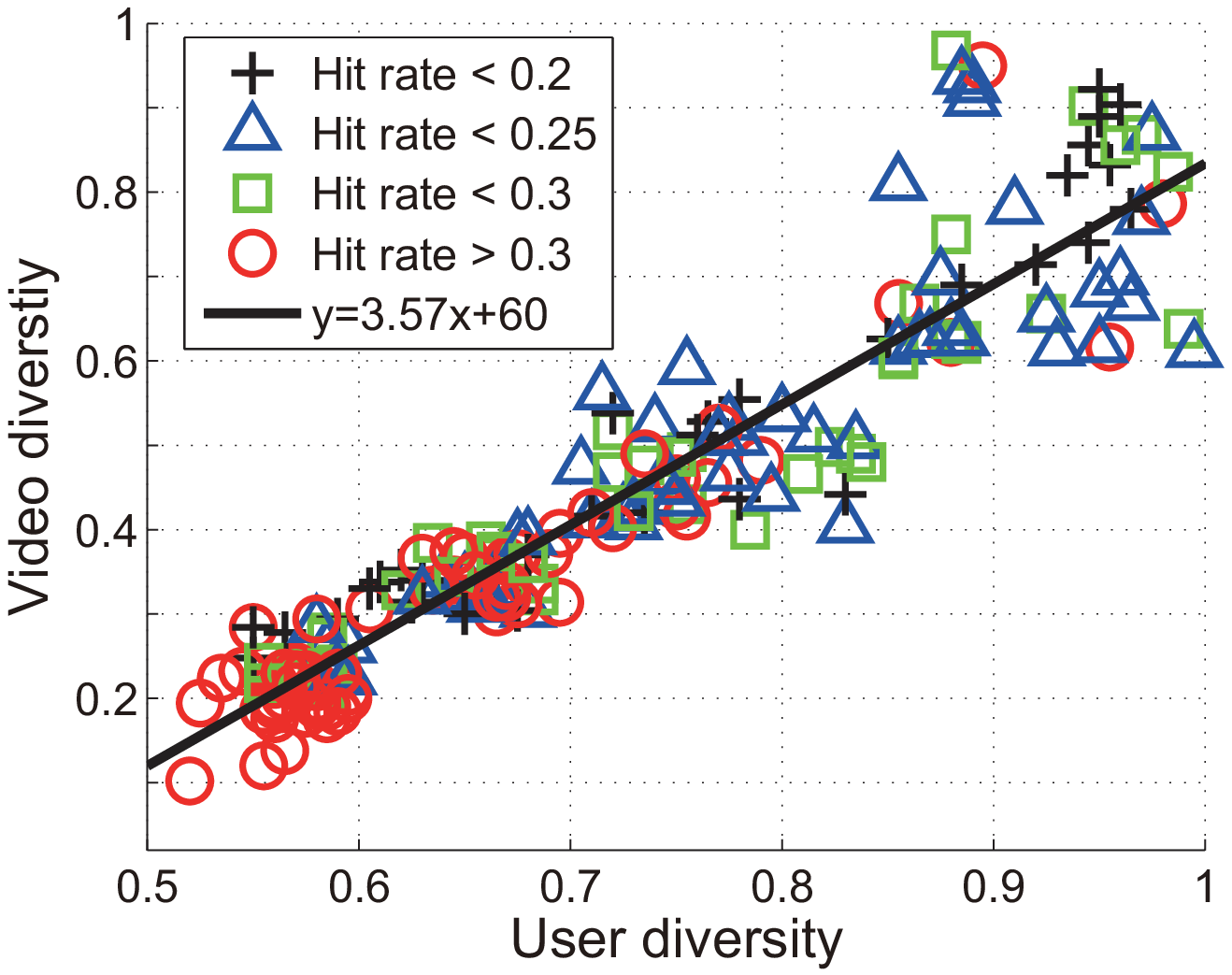}}
	\caption{Video and user diversity on edge network caching performance.}
	\label{fig:video_user_diversity}
\end{figure}

In previous measurement studies, we observe that mobile video request patterns exhibit both content and user diversity. We study the impact of such diversities on edge network content caching performance. We calculate the video diversity as the normalized number of unique videos requested in a location, and the user diversity as the normalized number of users in that location. Fig.~\ref{fig:video_user_diversity} shows the cache hit rates with different user diversity and video diversity for Wi-Fi caching and cellular caching, respectively. In Fig.~\ref{fig:diversity_BS}, we observe that \emph{lower video diversity and user diversity typically lead to higher cache hit rates}, because lower diversities lead to less unique content items requested. We observe different results in Fig.~\ref{fig:diversity_wifi}. For Wi-Fi caching, it is shown that the cache hit rate is considerably higher, and many samples with high cache hit rates appear with large user and video diversities. The result is also consistent with the measurement of location request entropy in Sec.~\ref{sec:entropy}. The different fitted lines imply that with the same user diversity, the Wi-Fi APs have larger video diversity on average. For caching strategies, \emph{network designers should deploy larger caches in locations with high video and user diversity to improve the quality of service}.

\subsubsection{Impact of User Mobility}

\begin{figure}[!t]
     \centering
         \subfigure[Caching]{
             \label{fig:mobility_BS}
             \includegraphics[width=0.47\linewidth]{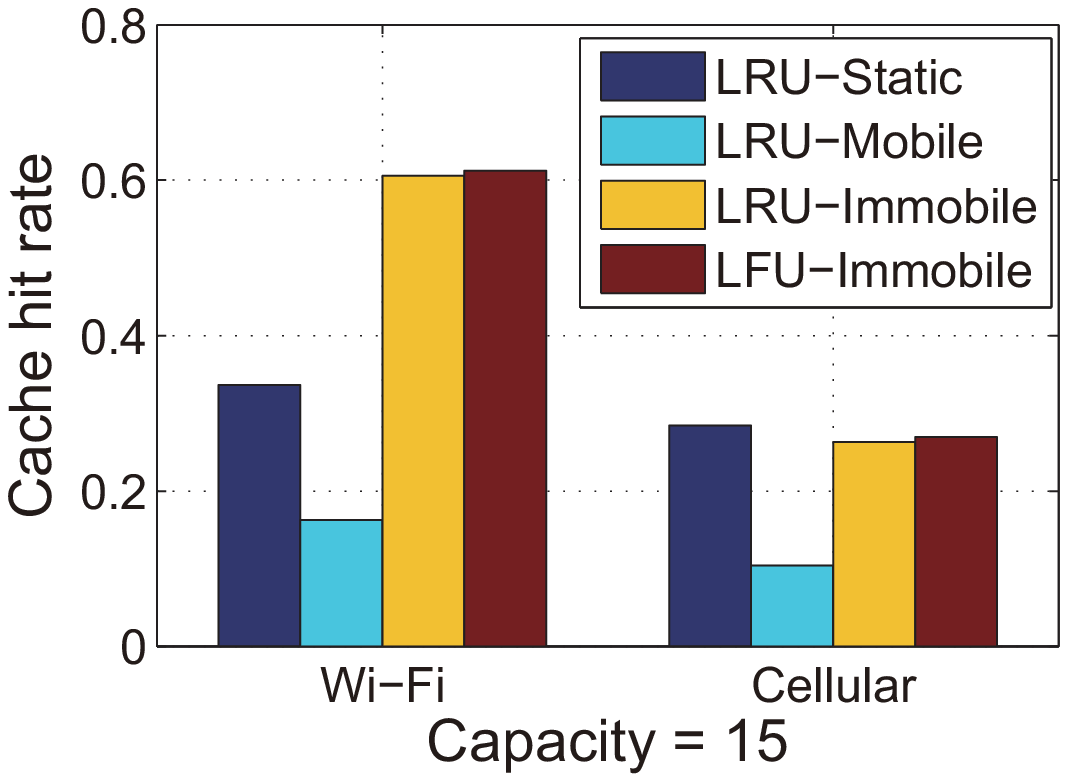}}
         \hfill             
         \subfigure[Jaccard similarity coefficient]{
             \label{fig:Jaccard}
             \includegraphics[width=0.47\linewidth]{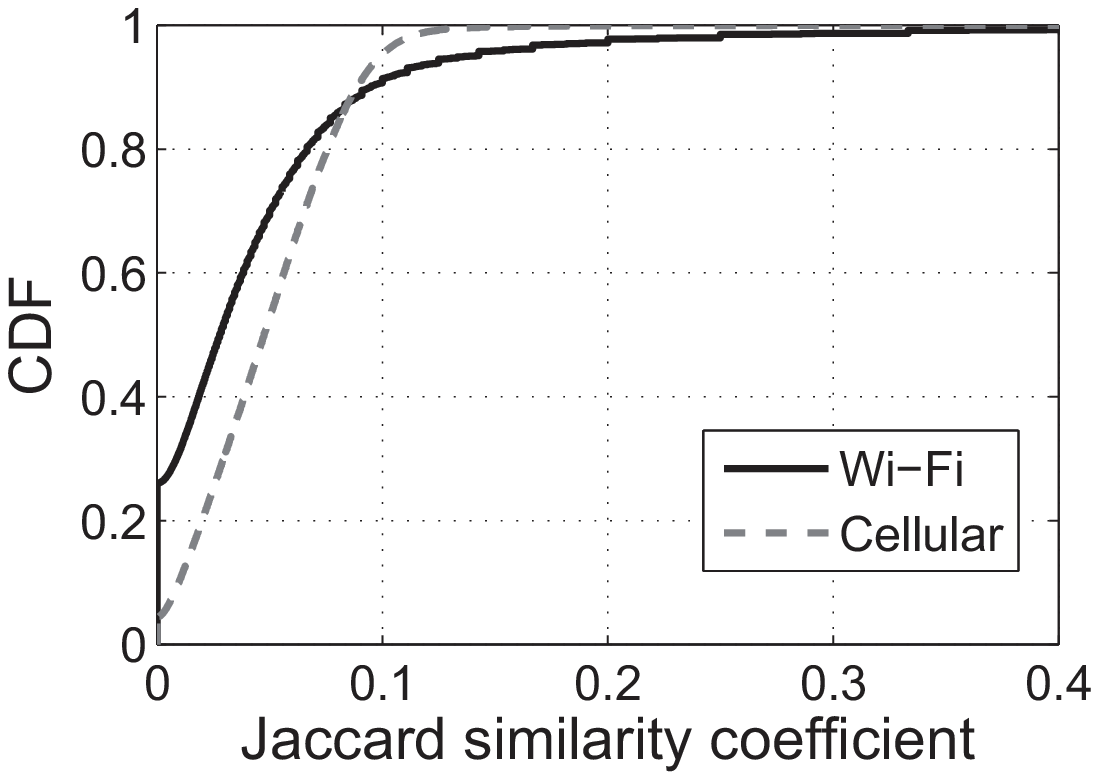}}
             
        \subfigure[Example]{
             \label{fig:user_example}
             \includegraphics[width=0.47\linewidth]{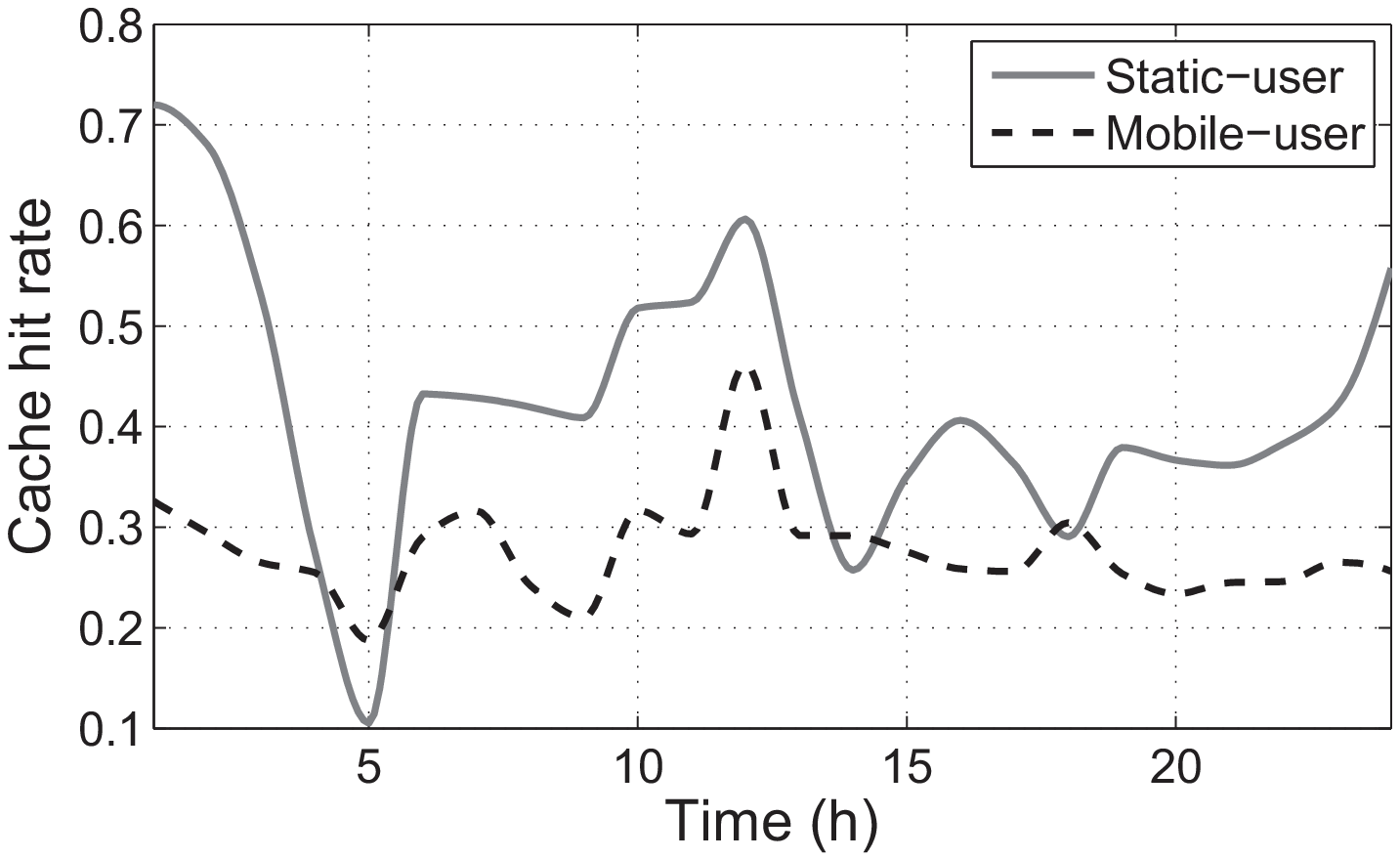}}
         \hfill             
         \subfigure[PoI locations]{
             \label{fig:PoI}
             \includegraphics[width=0.47\linewidth]{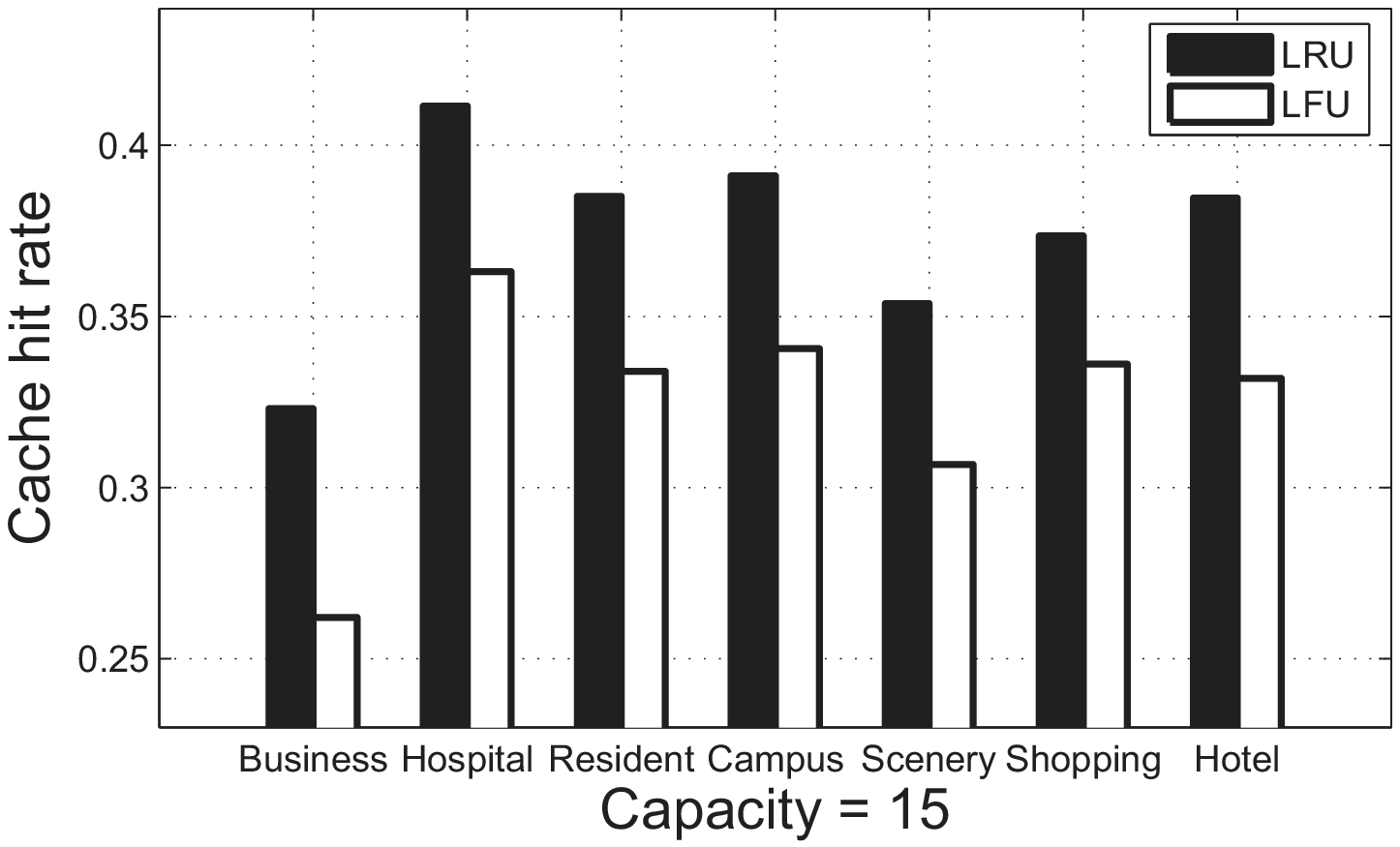}}                 
     \caption{Impact of user mobility.}
     \label{fig:mobility_cache}
 \end{figure}

We study the impact of user mobility on the edge network caching performance. In particular, we investigate the cache hit rates for multi-location users and single-location users. Fig.~\ref{fig:mobility_BS} shows that the cache hit rates of multi-location users are always lower than those of single-location users on both LRU and LFU. Thus, the user mobility has a highly negative influence on Wi-Fi/cellular. To determine the possible reasons for why multi-location users have considerably worse caching performance, we first compute the Jaccard similarity coefficient of the users' requested videos in start location $l_1$ and destination location $l_2$. The Jaccard similarity coefficient is 
$
J({l_1},{l_2}) = \frac{{\left| {S({l_1}) \cap S({l_2})} \right|}}{{\left| {S({l_1}) \cup S({l_2})} \right|}},
$
where $S({l_1})$ is a set consisting of the videos that users request in location $l_1$. The coefficient lies between $0$ and $1$, and the greater the value is, the more similarity they have.
Fig.~\ref{fig:Jaccard} depicts the CDF of the obtained similarity coefficients. The majority of location pairs have a similarity coefficient that is less than $0.4$, which indicates that the videos requested by users in different locations have greater differences. Second, we assume that the multi-location users are immobile and when moving to $l_2$ can still fetch content from $l_1$. The cache hit rates are recorded in Fig.~\ref{fig:mobility_BS}. Interestingly, the caching performance of Wi-Fi/Cellular is greatly improved, and LFU outperforms LRU, which is an opposite result of single-location users. Thus, \emph{the cache strategies based on LFU are more suitable for multi-location users}. The possible reason is that for destination location $l_2$, the user becomes a ``stranger''. Thus, $l_2$ has difficulty in satisfying the requests from $l_1$. As with LFU being better than LRU on multi-location users, the result verifies the measurement of location request entropy in Sec.~\ref{sec:entropy1}.
Fig.~\ref{fig:user_example} presents two instances of single-location and multi-location users. Fig.~\ref{fig:PoI} shows that the cache hit rates are different across different functional locations. This indicates that LRU and LFU have different cache hit rates across different locations. Furthermore, the caching performance gap between LRU and LFU is the smallest in shopping areas. One of the reasons is that there are many multi-location users in shopping area. This result also verifies the above observations that LFU is more suitable for multi-location users. 

\section{Cache Strategy Based on Measurement Insights}
\label{sec:application}

In this section, we design a geo-collaborative caching strategy for mobile video content delivery based on the measurement insights. We also compare its performance with conventional cache strategies.

\subsection{Caching Strategy}

Motivated by the measurement insights, we design a geo-collaborative caching strategy for mobile video delivery. Without loss of generality, we consider a general network architecture in which a set $\mathcal{L}$ of $L$ locations provide video content access to their users.

\emph{Cache storage.} For each location $l \in \mathcal{L}$, we divide the cache storage into $2$ parts: one is determined by users residing in the location (single-location users), and the other is determined by multi-location users requesting content there. According to the measurement results, the sizes of the two storage parts are determined by the fraction of the single-location users, i.e., a larger single-location user fraction indicates more storage for content to be requested by users residing in that location.

\emph{Cross-location reference.} According to our measurement studies in Sec.~\ref{sec:request_patterns}, locations with different functionalities have different request patterns. We propose a geo-collaborative caching strategy as follows. To enable content to be cached by cross-location reference, we propose a rank for locations using the information of user migrations: content requested in a location is referred more if there are more users migrating from/to that location, as follows. 
\begin{equation}
\label{eq:1}
\begin{array}{l}
    r_l^t = M \sum\limits_{i \in \mathcal{L}} o_{il}^{(t-W,t-1)} r_{i}^{t-1},
\end{array}
\end{equation}
where $o_{il}^{(t-W,t-1)}$ is the ratio of the users from location $i$ to $l$ over the total multi-location users in location $i$ in the previous time window $[t-W,t-1]$, $W$ is the time window (one day), and $M$ is a control parameter. 

\emph{Content to cache.} Let $\mathbf{x}_l^t$ denote the strategy to be applied for content replication in location $l$ in time slot $t$. An entry $x_{lv}^t = 1$ indicates that location $l$ will cache video $v$, and $x_{lv}^t = 0$ otherwise.
Similarly, $\mathbf{y}_l^t$ and $\mathbf{z}_l^t$ represent the caching strategies for single-location users in location $l$ and for multi-location users from other locations, respectively. For $\mathbf{z}_l^t$, we have
\begin{equation}
\label{eq:2}
\begin{array}{c}
\mathbf{z}_l^{t} =  \sum\limits_{i \in \mathcal{U}_l}  {\sum\limits_{j \in \mathcal{L}} r_j^t {d_{li}^{(t-W,t-1)} f_{ij}^{(t-W,t-1)} \mathbf{x}_{j}^{t-1}} },
\end{array}
\end{equation}
where 
$d_{li}^{(t-W,t-1)}$ is the fraction of request number of user $i$ over total request number in location $l$, $f_{ij}^{(t-W,t-1)}$ is the request distribution of user $i$ in location $j$,
and $\mathcal{U}_l$ is the set of users in location $l$. We iteratively calculate $\mathbf{z}_l^{t}$ in each time slot. 

Caching strategy $\mathbf{y}_l^t$ is determined by the popularity of videos requested by single-location users. For video $v$, its historical request number before time slot $t-1$ is $\rho_{v}^{t-1}$, and it is updated by $\rho_{v}^{t-1} = \rho_{v}^{(t-2,t-1)} + e^{-\mu} \rho_{v}^{t-2}$, where $\mu$ is a positive decay factor determined by the video category. To determine $\mathbf{y}_l^{t}$, location $l$ will cache videos requested with the largest $\rho_{v}^{t-1}$. Finally, $\mathbf{x}_l^t$ can be derived by the union of $\mathbf{y}_l^{t}$ and $\mathbf{z}_l^{t}$.

\subsection{Performance Evaluation}

We use the same simulator from the previous section to evaluate the cache strategy. In the experiments, to ensure the generality that each cellular base station (or Wi-Fi AP) has sufficient requests, only the top $10\%$ most requested cellular BSes (or Wi-Fi APs) are considered. 

\begin{figure}[!t]
	\centering
		\subfigure[Cache hit rate]{
			\label{fig:hit_rate}
			\includegraphics[width=0.47\linewidth]{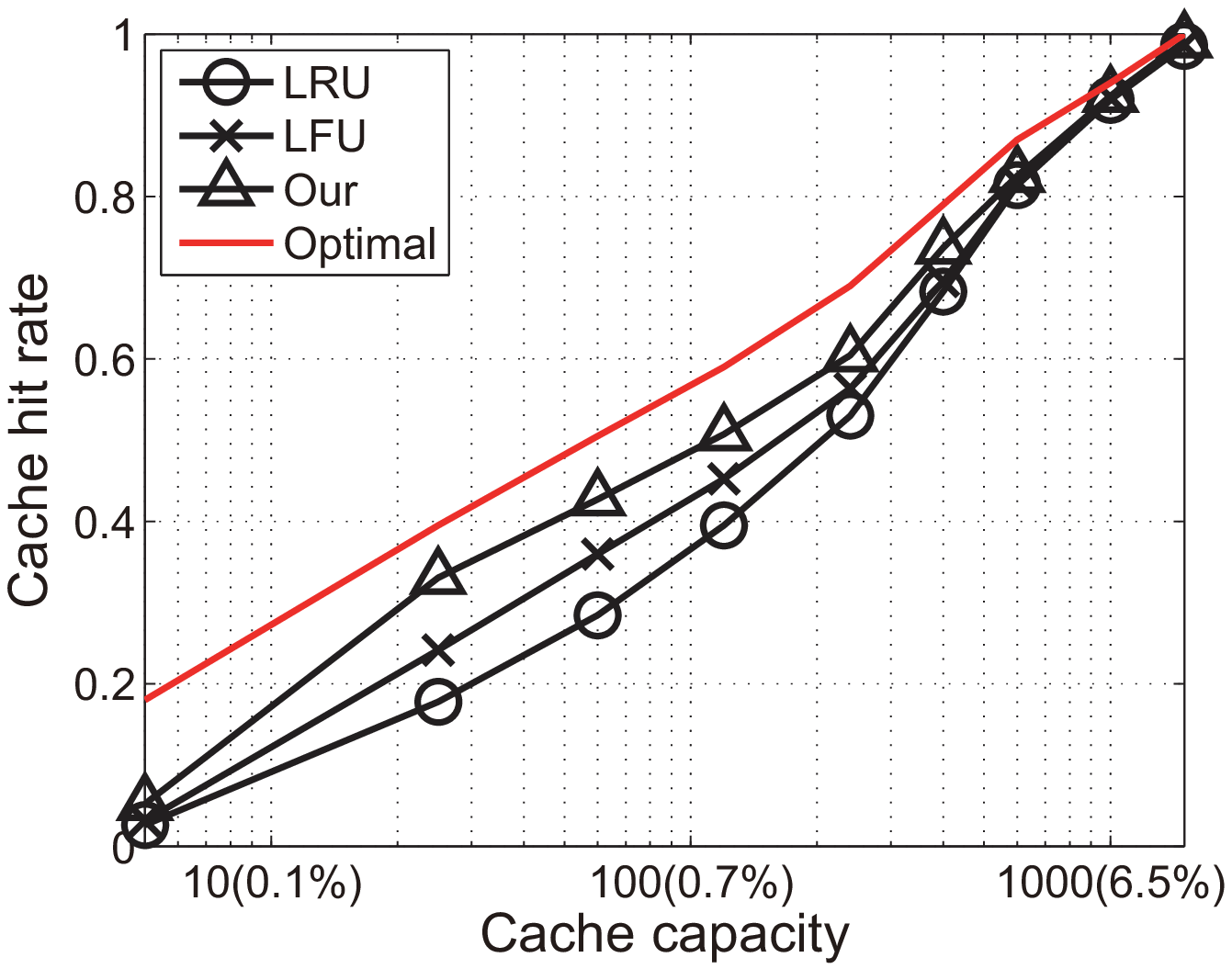}}
		\hfill
		\subfigure[Service rate]{
			\label{fig:service_rate}
			\includegraphics[width=0.47\linewidth]{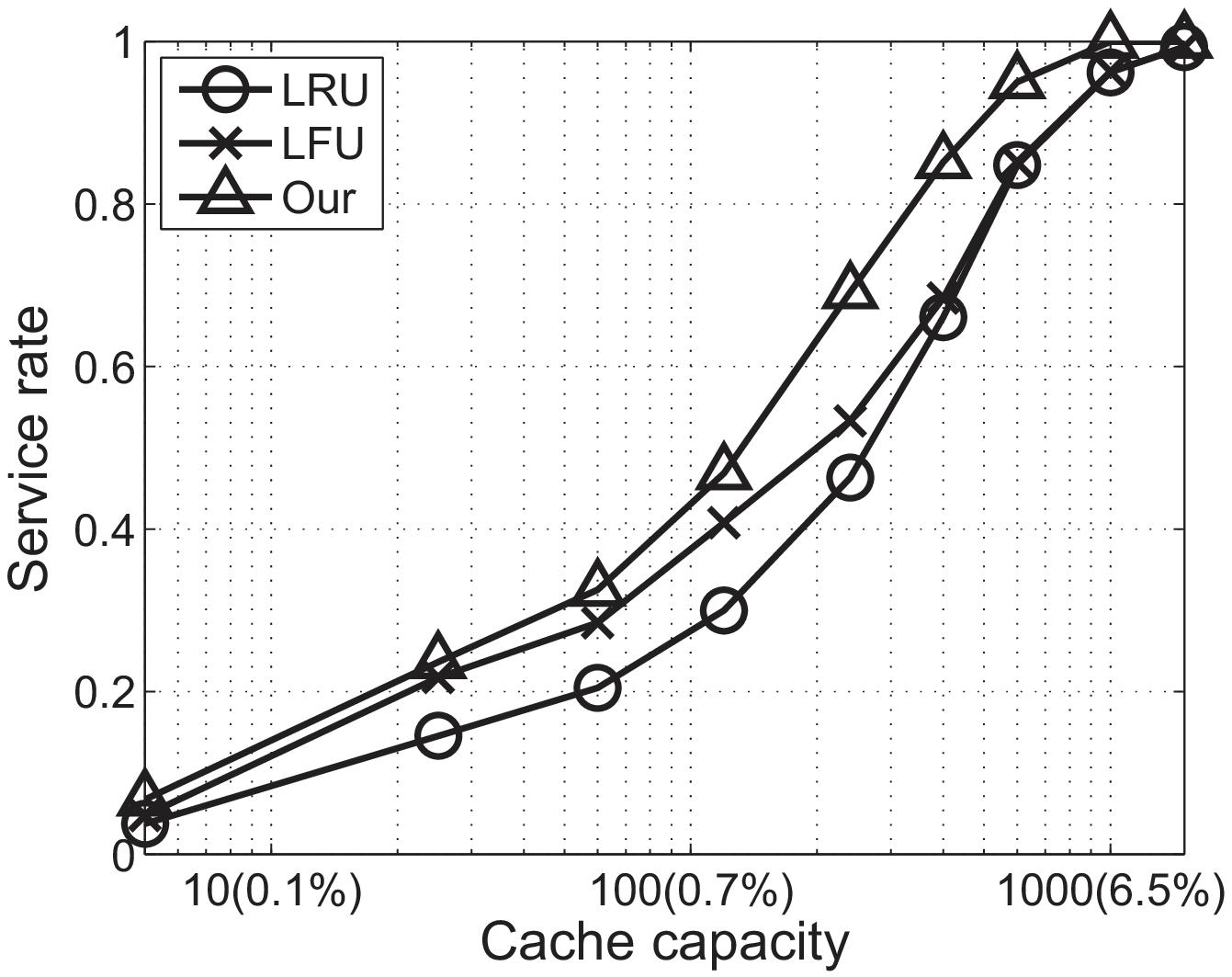}}
	\caption{Performance comparison (the percentage number in brackets is the ratio of cache capacity to the total number of content).}
	\label{fig:performance_comparison}
\end{figure}

We first study the impact of cache capacity on the cache hit rate. Fig.~\ref{fig:hit_rate} shows the cache hit rates of different caching strategies by varying the cache capacity from $1$ to $1500$. As expected, increasing the cache capacity increases the cache hit rate for all the caching strategies, as more requests are satisfied locally without requesting from the CDN servers. Compared with LRU and LFU, the gain of our method increases faster at the beginning. LRU, LFU and our strategy achieve similar cache hit rates when the cache capacity is large. The reason is that when the cache capacity is sufficiently large, each cellular BS can cache all the content and achieve a high cache hit rate.

Next, we study the \emph{service rate}, which is defined as the fraction of the number of users served by APs/BSes over the number of all users. Fig.~\ref{fig:service_rate} shows the service rates under different cache capacities. Compared with LRU and LFU, the gain of our strategy gradually increases as the cache capacity increases, indicating that the geo-collaborative cache strategy can potentially alleviate the original servers significantly. 

\section{Concluding Remarks} \label{sec:conclusion}

In this paper, we use measurement studies and trace-driven experiments to investigate the performance of edge network content caching for mobile video content delivery. We measure the spatial and temporal request patterns in mobile video systems and the user behaviors that have driven such request patterns. Our results show that the geographic request distribution in a mobile video system can be highly diverse, and the content requested varies among changing locations and periods. Such request patterns are generally determined by user mobility and preference behaviors, in which users exhibit regular commute behaviors, suggesting that joint caching strategies are promising for mobile video content delivery. Next, we compare the effectiveness of cellular and Wi-Fi based edge network caching solutions. Although Wi-Fi and cellular caching are promising, a number of factors including user mobility, content popularity, and cache capacity, have to be taken into consideration for edge network caching for mobile video delivery. Finally, we design a geo-collaborative caching strategy for mobile video delivery based on the measurement insights. Trace-driven experiments further verify the effectiveness of our design.

\bibliography{ref}
\bibliographystyle{IEEEbib}

\end{document}